\documentclass[apj]{emulateapj}
\slugcomment{{\sc Accepted:} May 5, 2008}
\usepackage{amsmath}
\usepackage{natbib}
\newcommand{\wsp}{$w_{sp}$}
\newcommand{\phip}{$\phi_{p}$}
\newcommand{\wspz}{$w_{sp}(z)$}
\newcommand{\phipz}{$\phi_{p}(z)$}
\newcommand{\dvdz}{$dV/dz\,d\Omega$}
\def\ltsim{\ {\raise-.5ex\hbox{$\buildrel<\over\sim$}}\ } 
\def\gtsim{\lower 0.5ex\hbox{$\; \buildrel > \over \sim \;$}}

\shorttitle{Calibrating Redshift Distributions}
\shortauthors{Newman}

\begin{document}


\title{Calibrating Redshift Distributions Beyond Spectroscopic Limits with Cross-Correlations}


\author{Jeffrey A. Newman \altaffilmark{1,2} }
\affil{University of Pittsburgh, Pittsburgh, PA 15260}
\email{janewman@pitt.edu}


\altaffiltext{1}{Hubble Fellow}
\altaffiltext{2}{present address: Department of Physics and Astronomy, University of Pittsburgh, 3941 O'Hara St., Pittsburgh, PA 15260}



\begin{abstract}

We describe a new method that can measure the true redshift
distribution of any set of objects that are studied only
photometrically.  Measuring the angular cross-correlation between
objects in the photometric sample with objects in some spectroscopic
sample as a function of the spectroscopic $z$, along with other,
standard correlation measurements, provides sufficient information to
reconstruct the redshift distribution of the photometric sample.  The
spectroscopic sample need not resemble the photometric sample in
galaxy properties, but must fall within its sky coverage. We test this
hybrid, photometric-spectroscopic cross-correlation technique with
Monte Carlo simulations based on realistic error estimates (including sample variance).  RMS errors in recovering both the mean redshift and $\sigma$ of the
redshift distribution for a single photometric redshift bin with true
distribution given by a Gaussian are $1.4\times
10^{-3}({\sigma_z/0.1})({\Sigma_p/10})^{-0.3}({dN_s/dz / 25,000
})^{-1/2}$, where $\sigma_z$ is the true Gaussian $\sigma$, $\Sigma_p$
is the surface density of the photometric sample in
galaxies/arcmin$^2$, and $dN_s/dz$ is the number of galaxies with a
spectroscopic redshift per unit $z$.  We test the impact of non-Gaussian redshift
outliers and of systematic errors due to
unaccounted-for bias evolution, errors in measuring autocorrelations,
photometric zero point variations, or mistaken cosmological
assumptions, and find that none will dominate measurement
uncertainties in reasonable scenarios.  The true redshift
distributions of even arbitrarily faint photometric samples may be
determined to the precision required by proposed dark energy
experiments ($\Delta\langle z \rangle \ltsim 3\times 10^{-3}$ at
$z\sim 1$) with this method.

\end{abstract}


\keywords{ galaxies: distances and redshifts, cosmology: large-scale structure of universe,  methods: miscellaneous,  surveys}



\section{Introduction}

Almost all cosmological tests require information about the distance or redshift of the objects studied.  For instance, the comoving length scale corresponding to baryon acoustic oscillations should remain fixed over time, but the corresponding angular size will depend on redshift in a cosmology-dependent manner; hence, if we measure this angular scale as a function of redshift, we may infer cosmological parameters \citep{2003ApJ...598..720S}.  Similarly, measuring weak lensing strength as a function of redshift can provide strong constraints on cosmological models \citep{1998ApJ...498...26K}, but the observed weak lensing signal will depend sensitively upon the redshift distribution of the background objects studied \citep{2002PhRvD..65f3001H}.  

However, although redshift ($z$) information is required for interpretation, it is infeasible to measure spectroscopic redshifts for the samples of hundreds of millions of extremely faint galaxies to be studied by proposed photometric dark energy probes such as the Large Synoptic Survey Telescope (LSST; \citealt{2001ASPC..232..347T,2005ASPC..339...95T}) or the Supernova / Acceleration Probe  (SNAP; \citealt{2000AAS...196.3212D,2004AAS...20513101P}), or even the millions of faint galaxies in samples now underway \citep{2006ApJ...647..116H}.   Hence, these projects will make use of photometric information to infer redshift distributions and to allow objects to be divided into multiple redshift bins for analysis.  This is possible because galaxy spectra are generally not featureless; as a spectrum redshifts through photometric passbands, its measured colors will vary with $z$ in ways that may be predicted based on spectroscopic observations of galaxies at similar redshifts or from the spectra of local analogues.  

However, these ``photometric redshifts" in general lack the precision of spectroscopically-derived redshifts, both because of photometric noise and outliers (e.g. in cases of overlapping galaxies) and because in some classes of galaxies (e.g. those forming stars most rapidly) the observable spectral features are weak at most redshifts (see, e.g., \citealt{2006A&A...457..841I}).  The true redshift distribution of objects with a given photometric redshift value may or may not be strongly or even singly-peaked, depending on the galaxy type, passbands used, photometric errors, etc.      

Because of these difficulties, dark energy experiments are unlikely to ever treat individual photometric redshifts as known with precision; instead, the approach taken in forecasts is to assume that objects will be divided into photometric redshift bins \citep{2006astro.ph..9591A}.  However, in order to obtain precision measurements of the properties of dark energy, both weak lensing and photometric baryonic acoustic oscillation (BAO) experiments require that the true redshift distribution of the objects in each bin be known with very high accuracy.  Projections for SNAP are that any overall bias in the mean redshift of a bin must be smaller than $2-4\times 10^{-3}$ in $z$ for dark energy constraints not to be degraded strongly \citep{2004ApJ...615..595H,2006ApJ...636...21M,2006MNRAS.366..101H}; for LSST, it is estimated that the mean redshift in each bin must be known to $\sim 2\times 10^{-3} (1+z)$ (\citealt{2006ApJ...644..663Z,2006JCAP...08..008Z,2006astro.ph..5536K,tysonconf},Tyson, Connolly, \& Newman, in prep.).  The true width of each bin must also be known, though with less precision ($\Delta \sigma_z < \sim 3\times 10^{-3} (1+z)$ for LSST, where $\sigma_z$ is the Gaussian sigma of the true redshift distribution; Tyson, Connolly, \& Newman, in prep.).   

These targets will be difficult to meet with standard spectroscopic techniques.  Both ongoing and proposed experiments reach depths far too faint (up to $R_{AB} \sim 30$ in the case of SNAP) for existing telescopes and spectrographs to measure redshifts.  Recent and ongoing surveys of faint galaxies using the largest telescopes available have obtained spectra for tens of thousands of galaxies to $R\sim 24$ or $I \sim 23$ (Davis et al. 2006, in prep; \citealt{2005A&A...439..845L}) and for a few thousand selected galaxies to $R\sim 25.5$ \citep{1999ApJ...519....1S,2003ApJ...592..728S,2004ApJ...604..534S}, but precise measurements of the redshift distributions of samples of galaxies with $R\sim 26-27$ or even fainter will be required in the future if proposed surveys are to reach their targets.  

Efforts to obtain true redshift distributions spectroscopically are made more difficult by the fact that faint galaxy surveys fail to obtain redshifts for a substantial fraction of their targets.  The DEEP2 Galaxy Redshift Survey, for instance, has obtained secure redshifts for $\sim 70\%$ of the galaxies studied \citep{2006MNRAS.370..198C}.  Roughly half of the missed targets appear to be star-forming galaxies at $z>1.4$ (which have no features within the DEEP2 spectral window) based on follow-up observations of blue DEEP2 redshift failures (C. Steidel, priv. comm.), but the redshift distribution of the remainder is unknown and currently being tested.  

Surveys of fainter galaxies have even lower success rates.  Despite integration times of more than thirty hours per object, the Gemini Deep Deep Survey (GDDS) only succeeded in measuring spectroscopic redshifts for $\sim 15\%$ of their targets with $24 < I< 24.5$, and $<50\%$ of objects with $23 < I < 24$ \citep{2004AJ....127.2455A}.  However, even with a completeness as high as the Sloan Digital Sky Survey ($\sim 99\%$; Schlegel et al. 2007, in prep.), if the objects missed are not a random subsample, redshift distributions could be biased beyond the tolerances of future dark energy surveys.  

Despite these difficulties, it is essential that there be some external method for testing photometric redshifts of faint galaxies, as it is quite likely that the spectral energy distributions (SEDs) of bright galaxies (whose redshifts are more easily obtained) should differ from those for fainter objects.  Both locally and at $z\sim 1$, the bluest galaxies (in rest-frame color) are intrinsically faint; they have no luminous analogues.  These issues make using SEDs from bright galaxies to determine photometric redshifts for fainter galaxies problematic.   Furthermore, at fainter magnitudes, higher-redshift galaxies will be more and more prevalent within a given photometric redshift bin.  These high-redshift galaxies may have contributions to their SEDs from metal-deficient ``Population III'' stars  that appear to have no local analogues \citep{2006Natur.440..501J}.  As an additional complication, Population III contributions should be greater in fainter, lower-mass galaxies than in more massive galaxies at all redshifts, given the evidence that lower-mass objects generally start forming stars later (e.g., Noeske et al., submitted). 

Despite these difficulties, if photometric dark energy surveys are to reach their goals, it is vital that we have some method of calibrating photometric redshifts with high precision \citep{2006astro.ph..9591A}.  In this paper, we describe a new method that can determine the true redshift distribution for any class of object (e.g. objects in a particular photometric redshift bin) by exploiting the fact that all galaxies at a given redshift cluster with each other.  We presume the existence of a large sample (or samples) of objects with spectroscopic redshifts.  The observed angular clustering between any two samples of galaxies will depend on both the intrinsic clustering of objects in the two samples with each other, and the degree to which they overlap in redshift (since clustering over extremely large distances is minimal).  Hence, by measuring the apparent angular cross-correlation between the positions of the photometric objects and the spectroscopic sample  {\it as a function of the spectroscopic z}, we may determine the actual redshift distribution of objects in the unknown class; the information provided by autocorrelation measurements for each sample allows us to break the degeneracy between correlation strength and redshift distribution.   

Similar cross-correlation techniques have been used in the past to measure correlation functions \citep{1985MNRAS.212..657P, 2006ApJ...644...54M} and luminosity functions \citep{1987MNRAS.227..115P}  at separations or depths where redshift surveys are incomplete; here, we explore their use to measure redshift distributions.  The angular cross-correlation can be measured with good precision in uniform, well-calibrated photometry, as required for future dark energy probes, as there will be many photometric galaxies near each spectroscopic galaxy on the sky.  The use of angular cross-correlations between photometric redshift bins to constrain the presence of redshift outliers has also been explored \citep{2006astro.ph..6098S,2006astro.ph..5302P}, but cannot determine redshift distributions in detail.  

The principal requirement of this method is that redshift survey data be available overlapping the photometric sample; however, the objects with redshifts need not be  similar to the target class (e.g., only high-confidence redshifts of relatively bright galaxies could be used when determining the redshift distribution of a sample of very faint galaxies).   In \S 2, we provide the theoretical underpinnings of this method.  In \S 3 we present Monte Carlo tests of its effectiveness for scenarios appropriate for current and future redshift surveys.  We evaluate potential sources of systematic error in \S 4, and in \S 5 we conclude.  Throughout this paper, we will use comoving coordinates for all distances and assume a cosmology with zero spatial curvature.  Where a specific cosmology must be adopted, we assume a flat $\Lambda CDM$ cosmology with matter density $\Omega_m$ = 0.3, dark energy density $\Omega_{\Lambda}=0.7$, and Hubble parameter $H_0=100h$ km s$^{-1}$ Mpc$^{-1}$.

\section{Measuring Redshift Distributions via Cross-Correlations}

\subsection{Basic Techniques}

\label{sec:basics}


Consider two sets of objects at cosmological distances, one with secure redshift measurements, which we will label '$s$' (for 'spectroscopic', though exceedingly accurate photometric redshifts might be used), and the other with unknown redshifts, which we will label '$p$' (for 'photometric').   In the most likely applications, the photometric sample would be a subset of objects in some photometric dataset -- e.g. objects in some bin of photometric redshift -- and the spectroscopic sample would result from one or more redshift surveys within the region of sky covered by $p$.  Although cross-correlation analyses require significant sample sizes, it should be possible to determine the bias and uncertainty in photometric redshifts as a function of $z$ by studying samples in a set of photometric-redshift bins.  

The mean comoving number density of objects in the photometric sample ($p$) a comoving real-space distance $r$ from an object in the spectroscopic sample ($s$) at redshift $z$, $n_p(r,z)$ will be
\begin{equation}
	\langle n_p(r,z) \rangle = n_p(z) (1 + \xi_{sp}(r,z) )\,,
\end{equation}
where $n_p(z)$ is the comoving number density of objects in sample $p$ at redshift $z$ and $\xi_{sp}(r,z)$ is the two-point cross-correlation function between samples $s$ and $p$.  Hence, $\xi_{sp}$ defines the excess probability of finding an object of class $p$ separated by a distance $r$ from an object of class $s$ that is at redshift $z$, above the probability if the two populations do not cluster with each other.  Finally, we  denote the probability distribution function for the true redshift of an object in the photometric sample by \phipz.

We assume that the surveyed objects are distant from us compared to the length over which correlations are significant and that $\xi_{sp}(r,z)$, $n_p(z)$, and \phipz\ may all be treated as constant over separations in the redshift direction comparable to that length (assumptions that all hold in typical high-redshift samples). In the distant-observer approximation, we may define $r^2 = \pi_l^2 +r_p^2 = \pi_l^2 + d_A(z)^2 \theta^2$, where $\pi_l$ is the comoving separation between two objects along the line-of-sight direction, $r_p$ is their projected comoving separation in the plane of the sky, $d_A(z)$ is the angular size distance to redshift $z$, and $\theta$ is their angular separation in radians.  We also define $l(z)$ to be the comoving distance to redshift $z$ given by $l(z)=\int_0^z c/H(z) dz$, where $c$ is the speed of light and $H(z)$ the Hubble expansion parameter at redshift $z$.  In calculations for measurements of angular correlations, we may ignore redshift-space distortions, so for a photometric object at redshift $z'$ separated by $\pi_l$ along the line-of-sight direction from a spectroscopic object at redshift $z$, $l(z') = l(z)+\pi_l$.  

The fundamental quantity we wish to recover is \phipz, the probability distribution for the true redshift of an object in $p$.  It can be related to $n_p$ and the mean surface density of objects in $p$ on the sky (in units of objects per steradian), denoted here by $\Sigma_p$:
\begin{eqnarray}
& \phi_p(z) & = \frac{dN_p}{dz\,d\Omega} \,/ \,\left(\int_0^{\infty} \frac{dN_p}{dz' \,d\Omega} \,dz' \right) \nonumber\\
&                & = n_p(z) \frac{dV}{dz\,d\Omega} \,/ \,\left(\int_0^{\infty} n_p(z') \frac{dV}{dz' \,d\Omega} \,dz' \right) \nonumber \\
&        	        & = n_p(z) \frac{dV}{dz\,d\Omega} \,/ \,\Sigma_p \nonumber \\
&        \label{eq:definephip}    &= \frac{n_p(z)}{\Sigma_p} d_A(z)^2 \frac{dl}{dz} ,  
\end{eqnarray} 
\\
\noindent where $dN_p / (dz\,d\Omega)$ is the number of objects in the photometric sample per unit redshift per steradian and $dV/(dz\,d\Omega)$ is the amount of comoving volume per unit redshift per steradian (equal to $d_A(z)^2 \,dl/dz$).

Equation 1 gives the excess number density of photometric objects near a spectroscopic object as a function of their real-space separation and the spectroscopic object's redshift.  However, what we are actually able to measure is the excess number of objects per unit area on the sky.   We therefore multiply Equation 1 by $dV/dz\,d\Omega$ and integrate over the possible redshifts of a photometric object, $z'$, to obtain $\langle \Sigma(\theta,z) \rangle$, the mean surface density of objects in $p$ an angle $\theta$ from an object in $s$ at redshift $z$.  We then obtain:
\begin{align}
 \langle \Sigma & (\theta,z) \rangle   =   \int_0^{\infty} n_p(z') d_A(z')^2 \frac{dl}{dz'} \,dz'  \nonumber\\
  & \hskip 0.7in {+ \int_0^{\infty} \,\xi_{sp}(r,z) n_p(z') d_A(z')^2 \,\frac{dl}{dz'} \,dz'} 	\nonumber \\
 & =   \int_0^{\infty} \Sigma_p\, \phi_p(z') \,dz' + \int_0^{\infty} \,\xi_{sp}(r,z)\, \Sigma_p\, \phi_p(z')  \,dz' \nonumber \\
 & =  \Sigma_p \, (1 + w_{sp} (\theta,z) \,)\,, \label{eq:definew}
\end{align}
\\
\noindent where $w_{sp}(\theta,z)$ defines the angular cross-correlation function between the spectroscopic and photometric samples and $\Sigma_p$ is the mean surface density of objects in $p$ over the sky.  
This constitutes the principal observable we will use to reconstruct the redshift distribution of the photometric sample (sample $p$).  


For convenience, we assume that all correlation functions may be described by power laws with linear biasing.  This assumption is somewhat unrealistic -- in real applications, precision measurements should use a halo model (cf. Cooray \& Sheth 2002 and references therein) or other more sophisticated methods -- but is sufficiently accurate to predict uncertainties for cross-correlation methods. 
If  $\xi_{sp}$ is represented by the power law form $\xi_{sp}(r) = (r / r_{0,sp})^{-\gamma}$, then the integrals in Equation \ref{eq:definew}b may be evaluated analytically to obtain:
\begin{equation}
w_{sp} (\theta,z) = \frac{\phi_p(z) \,H(\gamma)\, r_{0,sp}^\gamma \,\theta^{1-\gamma}\, d_A(z)^{1-\gamma}}{ {dl/dz}}\,,
		\label{eq:wsp-phip}
\end{equation}
where $H(\gamma)= \Gamma(1/2) \Gamma( (\gamma-1)/2 ) / \Gamma(\gamma/2)$ \citep{1980lssu.book.....P}, and we have treated \phipz\ and $d_A(z)$ as constant over the range in $z'$ for which $\xi_{sp}$ is nonnegligible (i.e., where $l(z)-l(z')$ is not much greater than $r_{0,sp}$).  

\nocite{2002PhR...372....1C}
\nocite{2005Msngr.121...42L} 

Although future applications of this method may make use of angular information (e.g. to constrain biasing models),  when determining the uncertainties resulting from application of cross-correlation methods below, we will focus on the integral of $w_{sp}$ within an angle equivalent to some comoving distance $r_{max}$ (which we will leave fixed with $z$) for simplicity.   We label this integrated \wsp\ ${\tilde w }(z)$, and define the angle corresponding to $r_{max}$ at a given $z$ to be $\theta_{max}(z)$.  Integrating Equation \ref{eq:wsp-phip} over $\theta$, we may relate $\phi_p$ to ${\tilde w }$ by the equation:
\begin{equation}
	\phi_p(z) = {\tilde{w}}(z) {3-\gamma \over 2 \pi} \frac{d_A(z)^2 \,dl/dz}{ H(\gamma)\, r_{0,sp}^{\gamma} \,r_{max}^{3-\gamma} } \, .
	\label{eq:phip}
\end{equation}

In general, a useful choice of $r_{max}$ should be large enough that the effects of nonlinear biasing (which can complicate modeling of correlation functions) are small, but small compared to the angular size of the photometric sample to minimize edge effects.  Our fiducial scenario will use $r_{max} =10\, h^{-1}$ Mpc, corresponding to roughly a quarter of a degree at $z=1$.  Correlation functions at both $z\sim 0$ and $z\sim 1$ are closely approximated by power laws at this scale \citep{2005ApJ...630....1Z,2006ApJ...644..671C}.

Inspection of Equations \ref{eq:wsp-phip} and \ref{eq:phip} shows that to determine $\phi_p(z)$ from $w_{sp}$ or ${\tilde{w}}$, we must know the basic cosmology (sufficient to determine $d_A(z)$ and $dl/dz$ modulo factors of $h$), $r_{0,sp}$, and $\gamma$.  We test the degree to which the cosmology must be known in \S \ref{sec:cosmo}; realistic uncertainities in cosmological parameters prove to have negligible impact.  

The same observations used to measure $w_{sp}$ provide sufficient information to determine $r_{0,sp}$ and $\gamma$ via the autocorrelation functions of the photometric and spectroscopic samples, $\xi_{pp}$ and $\xi_{ss}$.  This is true because, under our assumption of linear biasing, the cross-correlation $\xi_{sp}(r)$ must be given by the geometric mean of the autocorrelations of the two samples, $\xi_{sp}=(\xi_{ss} \times \xi_{pp})^{1/2}$; we will refer to this as the ``simple biasing" assumption hereafter.  This equation holds to high accuracy for the measured cross-correlations between subsamples in modern redshift surveys, even when their clustering differs strongly \citep{2007arXiv0708.0004C}.  The autocorrelation function of the spectroscopic sample, $\xi_{ss}$, is measurable directly from the spectroscopic sample, and is in fact a prime observable of redshift surveys; it thus remains only to determine $\xi_{pp}$.  

We may use the observed angular autocorrelation of the photometric sample, $w_{pp}(\theta)$, in conjunction with an initial guess for $\phi_p(z)$ to obtain the mean parameters of $\xi_{pp}$, since they are related by Limber's equation (evaluated for a power law correlation function, with scale length $r_{0,p}$ a function of $z$ but exponent $\gamma_p$ constant):
\begin{equation}
	w_{pp}(\theta) = H(\gamma_p) \,\theta^{1-\gamma_p} \int_0^{\infty} \phi_p^2(z)\, r_{0,p}(z)^{\gamma_p} \,\frac{d_A(z)^{1-\gamma_p}}{dl/dz} \, dz
\end{equation}
\\
\noindent \citep{1980lssu.book.....P}.  Note that $\gamma_p$ can be measured directly from the shape of $w_{pp}(\theta)$, so given a form for $\phi_p$ and the cosmology, the mean value of $r_{0,p}$ may be determined directly from its amplitude (in general, $\gamma$ varies only modestly -- $<10\%$ -- even amongst samples of galaxies with very different biasing; \citealt{2005ApJ...630....1Z}).  This procedure may be iterated by using the derived parameters of $\xi_{pp}$ to determine $\xi_{sp}$ and hence $\phi_p$ from cross-correlations, then redetermining $\xi_{pp}$ using this $\phi_p$, and then refining $\phi_p$ from cross-correlations, etc. until convergence is reached.  Although $w_{pp}$ yields only a weighted mean of the value of $r_{0,p}(z)$, the consequences of feasible amounts of variation in the bias of the photometric sample with redshift are modest; we demonstrate this in \S \ref{sec:bias}.

To summarize: the most basic large-scale structure measurements possible where a photometric sample overlaps a spectroscopic one -- the two-point autocorrelation functions of each sample with itself, and their cross-correlation on the sky, measured as a function of spectroscopic redshift -- provide sufficient information to reconstruct the redshift distribution of the photometric sample.  In the remainder of this paper, we will attempt to determine the uncertainties, both random and systematic, that should result from applying such methods to realistic samples.

\subsection{Error Estimates}

\label{sec:errors}

We begin by estimating the error in a measurement of $\phi_p(z)$ in some small bin of redshift of centered at $z$ and of width $\Delta z$.   We presume here that the errors in $\phi_p$ will be dominated by the uncertainty due to counting statistics in a measurement of ${\tilde w}$,
the integral of $w_{sp}$ within the angle $\theta_{max}(z)$.  Poisson uncertainties should dominate when ${\tilde w }$ is small (the ``weak-clustering" limit), which should always be the case unless $\phi_p$ is unrealistically narrow  \citep{1980lssu.book.....P}.  We defer investigation of possible systematic errors to \S \ref{sec:systematics}. 

Modulo the modest impact of sample variance (see \S \ref{sec:cv}), we expect uncertainties to be dominated by errors in ${\tilde w}$, as the autocorrelations $\xi_{ss}$ and $w_{pp}$ should be measured more precisely. Then, applying standard propagation of errors to Equation \ref{eq:phip}, $\sigma(\phi_p(z)) / \phi_p(z)= \sigma({{\tilde{w}}}) / {\tilde w }$. Furthermore, $\sigma({{\tilde{w}}}) / {\tilde w }$ must equal the uncertainty in the total excess (over random) number of neighbors in $p$ surrounding any member of $s$ due to clustering (which we will label $\sigma(N_c)$), divided by the expected number of these neighbors (denoted by $N_c$), as ${\tilde w }$ is directly proportional to $N_c$ by definition.   Since these quantities are simple to predict, we will determine $\sigma(\phi_p(z)) / \phi_p(z)$ by calculating the equivalent quantity, $\sigma(N_c) / {N_c}$.  


In the weak-clustering limit (i.e., so long as ${\tilde w}$ is small, as is true here), the uncertainty in $N_c$, $\sigma(N_c)$, is given simply by the Poisson uncertainty in the expected total number of spectroscopic-photometric pairs if there is no clustering  \citep{1980lssu.book.....P}.    Thus,
 \begin{eqnarray}
&	\sigma^2(N_c) &= \left( \Sigma_p \,\pi \theta_{max}^2 \right) \left( \frac{dN_s}{dz} \Delta z\right) \nonumber\\
&				&=   \pi\, \Sigma_p \,\frac{dN_s}{dz} \,\Delta z \,\left(\frac{r_{max}}{d_A(z)}\right)^2,   	\label{eq:signca}
	\label{eq:signc}
\end{eqnarray}
where $dN_s/dz$ gives the actual redshift distribution of the spectroscopic sample.  Inside the first parentheses in Equation \ref{eq:signca} is found the expected number of members of the photometric sample within $\theta_{max}$ of each object in the spectroscopic sample, while inside the second parentheses we give the number of objects in $s$ within the designated redshift bin.  

We may determine $N_c$, the total number of excess spectroscopic-photometric ($s-p$) pairs within separation $\theta_{max}$ over random due to correlations, by integrating the real-space two-point cross-correlation function over the relevant volume:

\begin{align}
N_c &= \left(\int_0^{r_{max}} 2\pi r_p \int_0^{\infty} n_p(z') \,\xi_{sp}(r,z) \,dz'  dr_p \right)\times \left(\frac{dN_s}{dz} \,\Delta z\right) \nonumber\\
&= \left(\int_0^{r_{max}} 2\pi r_p \int_0^{\infty} {\phi_p(z') \Sigma_p \over d_A(z')^2 \,dl/dz'}  \,\xi_{sp}(r,z) \,dz'  dr_p \right) \nonumber\\
& {\hskip 0.4in \times \left(\frac{dN_s}{dz} \,\Delta z\right)} \nonumber\\
&= 2 \pi H(\gamma) {\phi_p(z)\, \Sigma_p \over d_A(z)^2 \,dl/dz} r_{0,sp}^\gamma\, \frac{dN_s}{dz} \,\Delta z \  \int_0^{r_{max}}  r_p^{2-\gamma} \,dr_p  \nonumber\\
&= {2 \pi H(\gamma)\over 3-\gamma}  {\phi_p(z) \,\Sigma_p \over d_A(z)^2 \,dl/dz} r_{0,sp}^\gamma \, r_{max}^{3-\gamma} \,\frac{dN_s}{dz} \,\Delta z \label{eq:nc}
\end{align}
\\
where again we have separated the number of members of the photometric sample around each member of the spectroscopic sample from the total number of members of the spectroscopic sample within $\Delta z$ using parentheses before combining them, and assumed that \phipz\ is approximately constant over the range in $l(z')$ where $\xi_{sp}(r,z)$ is nonngeligible.  

Combining Equations \ref{eq:signc} and \ref{eq:nc}, we then find:
\begin{align}
\sigma(& \phi_p(z))  = \phi_p(z) \frac{\sigma(N_c)}{N_c}  \label{eq:errors} \\
& = {3 - \gamma \over 2 \sqrt{\pi} H(\gamma)} \left(\Sigma_p\, \frac{dN_s}{dz} \,\Delta z\right)^{-1/2} \,{d_A(z) \,dl/dz \over r_0^\gamma \, r_{max}^{2-\gamma}} \, . \nonumber
\end{align}

It is worth noting that these errors scale only very slowly with $r_{max}$, since for typical galaxy samples $\gamma \approx 1.6-1.9$.  As a consequence, it is possible to minimize nonlinear effects not only by measuring correlations to large separations (i.e., increasing $r_{max}$), but also by excluding the smallest separations ($r_p <1-2 h^{-1}$ Mpc) from the calculation of the integrated correlation, ${\tilde w }(z)$.  Predicted measurement uncertainties increase only modestly if small-separation pairs are not considered; for $r_{max}=10 h^{-1}$ Mpc, excluding the central $2 h^{-1}$ Mpc increases overall errors by roughly 15\%.  That exclusion radius is larger than the maximum $r_p$ where non-power law cross-correlations have been observed in studies of the clustering of blue, starforming galaxies about galaxy groups \citep{2006ApJ...638..668C}, which is likely to be a maximally pathological case.


\subsection{Sample Variance} 

\label{sec:cvtheory}

These error estimates have ignored the fact that the mean density of the regions where we perform these cross-correlation measurements at a given $z$ may be higher or lower than the Universal mean, the effect commonly referred to as ``sample" or ``cosmic" variance.  Thus, the recovered redshift distribution of the photometric sample, $\phi_r(z)$, will differ from the distribution that would be obtained if an infinite volume were surveyed, \phipz.  If the region over which cross-correlations are measured corresponds to the full area covered by the photometric sample, then in fact $\phi_r$ may be the desired quantity, rather than the ``true", underlying distribution, \phip.  Outside of this regime, we must consider the impact that sample variance will have on our recovery of \phipz.

In particular, we can place two limits on the impact of sample variance.  If cross-correlations are measured over very large areas of sky (hundreds of square degrees), sample variance should be negligible compared to other sources of error, and the random errors in the recovery of \phipz\ will simply be given by Equation \ref{eq:errors}.  We will assess how much area is sufficient to reach this regime in \S \ref{sec:cv}.  Thus, our previous error estimates are in fact a lower limit on measurement uncertainties from cross-correlation techniques.

If only a few fields with small areas are surveyed, the impact of sample variance is much greater.  However, the spectroscopic sample may be used to limit this impact, as the variations in density will cause proportional variations in the number of galaxies found in a given redshift bin (compared to a smooth model).  Since \wspz\ measures the excess number of companions per spectroscopic object at a given $z$, these variations in the redshift distribution of the spectroscopic sample, $s$, do not affect the measured \phipz\ directly.  However, there will be corresponding variations in the number of members of the photometric sample $p$ at that redshift, with the amplitude of those variations proportional to the ratio of the large-scale bias of sample $p$ to that of sample $s$.  

Since that ratio of biases would be determined in the process of measuring \phipz\ from cross-correlations, we may estimate the universal value of \phipz\ from the value reconstructed in a particular region of the sky:
\begin{equation}
	\phi_p(z) = {\phi_r(z) \over 1+ b_p/b_s \, \Delta_s(z)}\,,
\end{equation}
where $b_p$ and $b_s$ are the linear, large scale biases of the photometric and spectroscopic samples ($p$ and $s$, respectively) and $\Delta_s(z)$ is the fractional deviation of $dN_s/dz$ at a given redshift from a smooth model; i.e., $[ (dN_s/dz)_{observed} - (dN_s/dz)_{true}]/(dN_s/dz_{true})$.  Because $(dN_s/dz)_{observed}$ is determined from the finite number of spectroscopic objects within $\Delta z$, it will be subject to Poisson variance; hence $\Delta_s(z)$ has a measurement uncertainty $\sigma(\Delta_s) = (dN_s/dz \times \Delta z)^{-1/2}$.  This will propagate into a residual uncertainty in \phipz\ of $(b_p/b_s) (dN_s/dz \times \Delta z)^{-1/2} \phi_p(z)$ (taking $(1+(b_p/b_s) \Delta_s)^2 \sim 1$, which holds for all realistic survey characteristics).  This error is independent of counting-statistics errors; thus when assessing the maximal impact of sample variance, we combine it with the measurement uncertainty given by given by Equation \ref{eq:errors} following standard propagation of errors.  

Because it increases overall uncertainties the most where \phipz\ is largest, the net effect of sample variance after correcting with $(dN_s/dz)_{observed}$ is to reduce modestly the advantages of samples with tight redshift distributions or high surface densities.  Since errors from both counting statistics and sample variance scale as $(dN_s/dz \times \Delta z)^{-1/2}$, though, both of these sources of uncertainty will be reduced by the same fraction if  $dN_s/dz$ is increased.  In our Monte Carlo simulations, we assume $b_p = b_s$.  We expect that for typical datasets $b_p < b_s$, as photometric samples should go fainter than spectroscopic samples, and fainter objects tend to have lower bias \citep{2005ApJ...630....1Z,2006ApJ...644..671C}, making this assumption an upper limit.  In the simulations below, we will estimate the errors in the reconstruction of \phipz\ both when sample variance is negligible and when $dN_s/dz$ is used for corrections, in order to bracket the possibilities.


\section{Monte Carlo Tests}

\subsection{Basic Scenarios}

\label{sec:results}

We now investigate the degree to which cross-correlation techniques can recover true redshift distributions for photometric samples.  For our most basic scenario, we adopt a simple \phipz\ distribution given by a Gaussian with mean redshift $z_0$ (which we generally take to be 1, near the peak of sensitivity of most dark energy measurement methods) and standard deviation $\sigma_z$; i.e., 
\begin{equation}
\phi_p(z) = g(z) = {1 \over \sqrt{2 \pi} \sigma_z} \exp{(-{(z-z_0)^2 \over 2\sigma_z^2})} \, .
\label{eq:gz}
\end{equation}
\vskip 0.1in\noindent We then use Monte Carlo techniques to test the recovery of both the mean and standard deviation of this distribution given a spectroscopic sample with some redshift distribution $dN_s/dz$.  A test of these techniques with catalogs taken from an N-body simulation is now underway, and finding similar results (Wittman 2008, in prep.). 

\begin{figure}
\epsscale{1.20}
{\hskip -0.15in \plotone{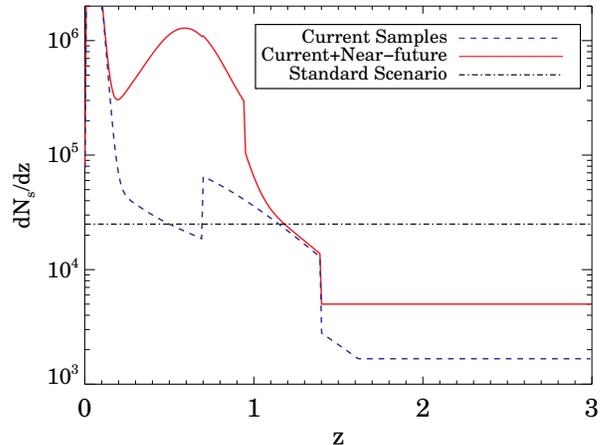} }
\caption{Redshift distributions assumed for current spectroscopic samples (blue dashed line) and future samples (red solid line).  The assumed characteristics of each sample are given in Table \ref{tab:dndz}.  The differences are the addition of an intermediate-redshift survey, PRIMUS (Eisenstein et al. 2007, in prep.); a baryonic oscillation survey, WiggleZ (Glazebrook et al. 2007, in prep.); zCOSMOS \citep{2005Msngr.121...42L}; and larger samples at $z>2$ in the near-future scenario.  These samples were used to produce the Monte Carlo realizations shown in Figure \ref{fig:montecarlo}.  The black, dot-dashed line indicates the assumption used for our standard scaling scenario, which approximates current redshift samples at $z\sim 1$.  \label{fig:dndz}}
\end{figure}

\begin{figure*}
\epsscale{1.15}
\plottwo{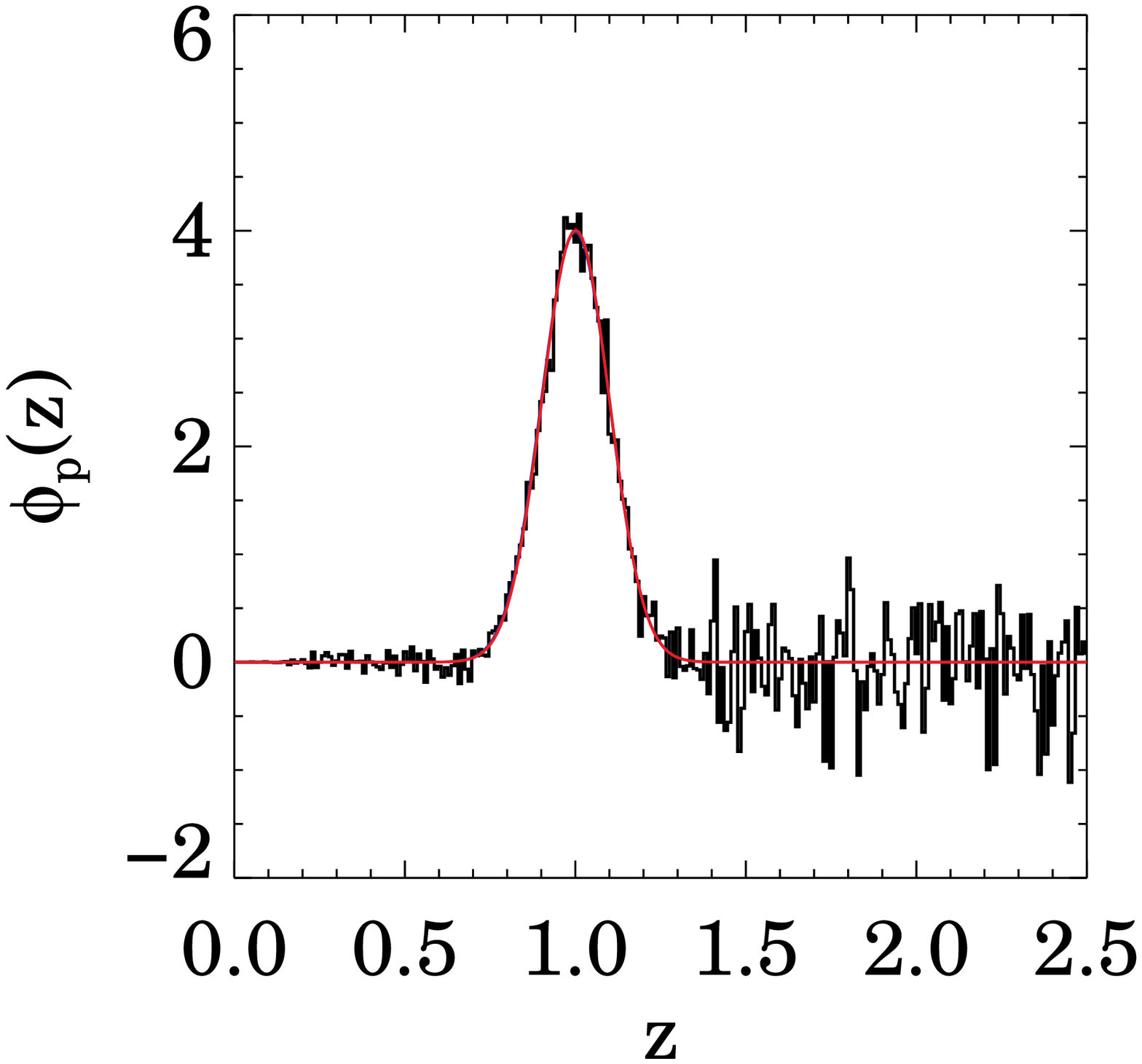}{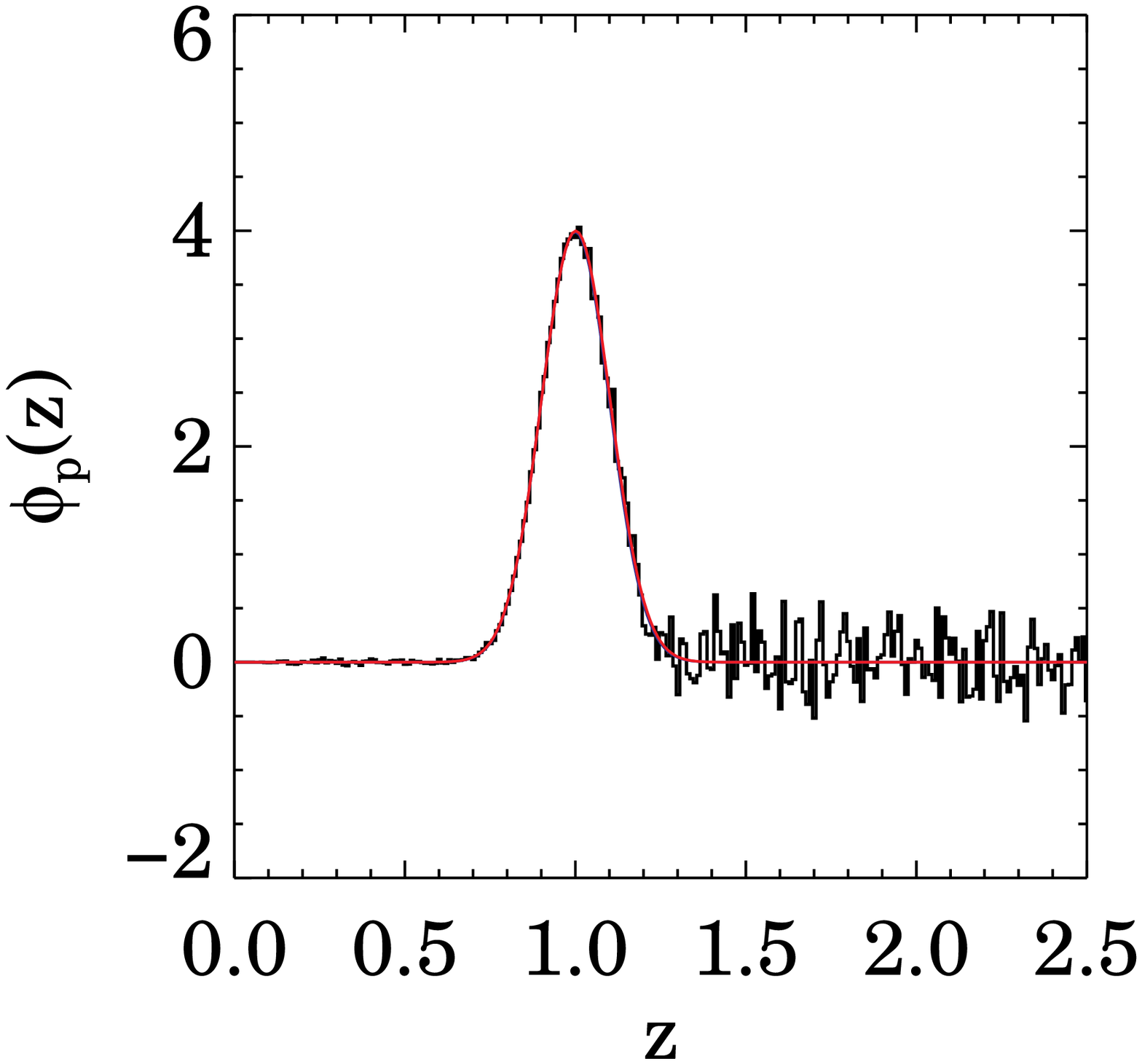}
\caption{Examples of individual Monte Carlo realizations for the recovery of \phipz\ using the combinations of current spectroscopic datasets (left) or of current and future datasets (right) shown in Fig. \ref{fig:dndz}.   Each realization was generated by randomly drawing from realistic error distributions for the recovery of $\phi_p(z)$ in bins of width $\Delta z=0.01$.  Plotted in blue is the true, input redshift distribution, given by Equation \ref{eq:gz} with $\sigma_z=0.1$.  The black histogram shows one realization for the distribution measured using cross-correlation techniques, with realistic errors determined as described in \S \ref{sec:errors}.  Shown in red is the distribution determined from a least-squares fit to the simulated data shown by the black histogram.  The recovery is good enough in each case that the blue curve is essentially invisible. \label{fig:montecarlo}}
\vskip 0.4in
\end{figure*}

To perform these Monte Carlo tests, we generate realizations of the recovered \phipz\ in a large number of bins of width $\Delta z$, adding to the true \phipz\ in each bin an error drawn randomly from a Gaussian distribution with mean zero and standard deviation given by $\sigma$(\phipz) for that bin, incorporating both counting statistics and sample variance as described above.  Where $\sigma_z > 0.1$, we use bins of width $\Delta z=0.01$; otherwise, the bin width used is $\Delta z=0.01 \times (\sigma_z/0.1)$ to ensure the peak is well-resolved.  For each realization, we fit for the parameters of \phip\ with standard nonlinear least-squares techniques.  To ensure stability in the fitting, we provide initial guesses for the parameters given by the true value plus a random value drawn from a Gaussian distribution with standard deviation 10\% of the true value.  This is effectively equivalent to assuming that the true distribution parameters are known to 10\%, far worse than the tolerances for most dark energy experiments and much larger than the errors resulting from the cross-correlation measurements.  We show example realizations and reconstructions (based on the redshift distributions shown in Figure \ref{fig:dndz}) in Figure \ref{fig:montecarlo}.

\begin{deluxetable*}{lccc}
\tabletypesize{\scriptsize}
\tablecaption{Assumed Redshift Survey Samples}
\tablewidth{0pt}
\tablehead{\colhead{Survey name} & \colhead{\# of high-confidence{\it z}'s} & \colhead{$z_0$\tablenotemark{a} or redshift range} & \colhead{Reference} }
\startdata
\cutinhead{Current Samples}
Sloan Digital Sky Survey (SDSS) & 800,000 & 0.017 & \citealt{2002AJ....124.1810S} \\
AGN \& Galaxy Evolution Survey (AGES) & 10,000 & 0.09 & \citealt{2004AAS...205.9402K} \\
DEEP2 Galaxy Redshift Survey, EGS & 8,200 & 0.225 & Davis et al. 2006 \\
DEEP2 Galaxy Redshift Survey, non-EGS & 17,000\tablenotemark{b} & 0.225\tablenotemark{b} & Faber et al. 2007 \\
VIMOS/VLT Deep Survey & 10,000\tablenotemark{c}  & 0.27 &  {Le F{\`e}vre} et al. 2005\\
Lyman/Balmer break samples, $1.5<z<4$ & 2,500 & $1.5<z<4$ & \citealt{1999ApJ...519....1S,2003ApJ...592..728S,2004ApJ...604..534S} \\
\cutinhead{Near-future Samples}
WiggleZ & 350,000 & $0.25 < z < 1$\tablenotemark{d} & Glazebrook et al. 2007, in prep. \\
PRIMUS & 300,000 & 0.23\tablenotemark{e} & Eisenstein et al. 2007, in prep. \\
zCOSMOS, $I<22.5$ & 5,000\tablenotemark{c} & 0.23 & Lilly et al. 2006 \\
zCOSMOS, high-z & 2,500\tablenotemark{c} & $1.5 < z < 2.5$ & Lilly et al. 2006 \\
Lyman/Balmer-break samples, $1.5<z<4$ & 5,000 & $1.5<z<4$ & N/A \\

\enddata
\tablenotetext{a}{Except where redshift ranges are specified, we assume redshift distributions are proportional to $z^2 e^{-z/z_0}$, and so have median redshift given by $2.67 z_0$ and mean redshift $3 z_0$.  We then use published median redshifts or photometric depths to estimate $z_0$, as described in the text.}
\tablenotetext{b}{Outside of the Extended Groth Strip (EGS), DEEP2 uses a color cut complete for $z>0.75$ ($\sim 50\%$ complete at $z=0.7$).  [OII] 3727 and the 4000-\AA break leave the DEIMOS spectral window at $z \gtsim 1.4$, so redshift success is minimal beyond that point.  We therefore include only redshift quality=4 objects with $0.7 < z < 1.4$ in this count and the model redshift distribution.  }
\tablenotetext{c}{For VVDS and zCOSMOS, we take a 20\% rate of flag=4 redshifts, following \citet{2005A&A...439..845L}, and take a full VVDS sample size of 50000 objects.   We optimistically assume that the median redshifts given in \citet{2005A&A...439..845L} apply for the flag=4 galaxies, though very few of them are at $z>1$ \citep{2005A&A...439..863I}.} 
\tablenotetext{d}{The redshift distribution from test WiggleZ observations may be roughly approximated by a Gaussian of mean 0.55 and $\sigma$ 0.25, truncated at $z=0.25$ and $z=1$.} 
\tablenotetext{e}{We presume that, in addition to being $I$-band limited, PRIMUS will generally not measure redshifts for objects with $z>0.95$ (A. Coil, priv. comm.).} 
\label{tab:dndz}
\end{deluxetable*}


For every scenario tested in this paper, we generate ten thousand realizations of this sort, measuring the parameters of \phipz\ each time; we then determine the mean and standard deviation of the results for each parameter to test the efficacy of cross-correlation methods.  For our basic scenarios, we vary three things: the width of \phipz, $\sigma_z$; the surface density of members of the photometric sample on the sky, $\Sigma_p$; and the redshift distribution of the spectroscopic sample, $dN_s/dz$.   

We here ignore the weak cross-correlations induced by gravitational lensing.  These correlations can be predicted directly from the observed galaxy number counts \citep{2005ApJ...633..589S}.  Alternately, it should be possible to iteratively remove the lensing-induced signal once we have an estimate of \phipz\, as that will allow us to predict how much cross-correlation with members of $s$ at a given $z$ should result from lensing.



\begin{figure}
\epsscale{1.20}
\plotone{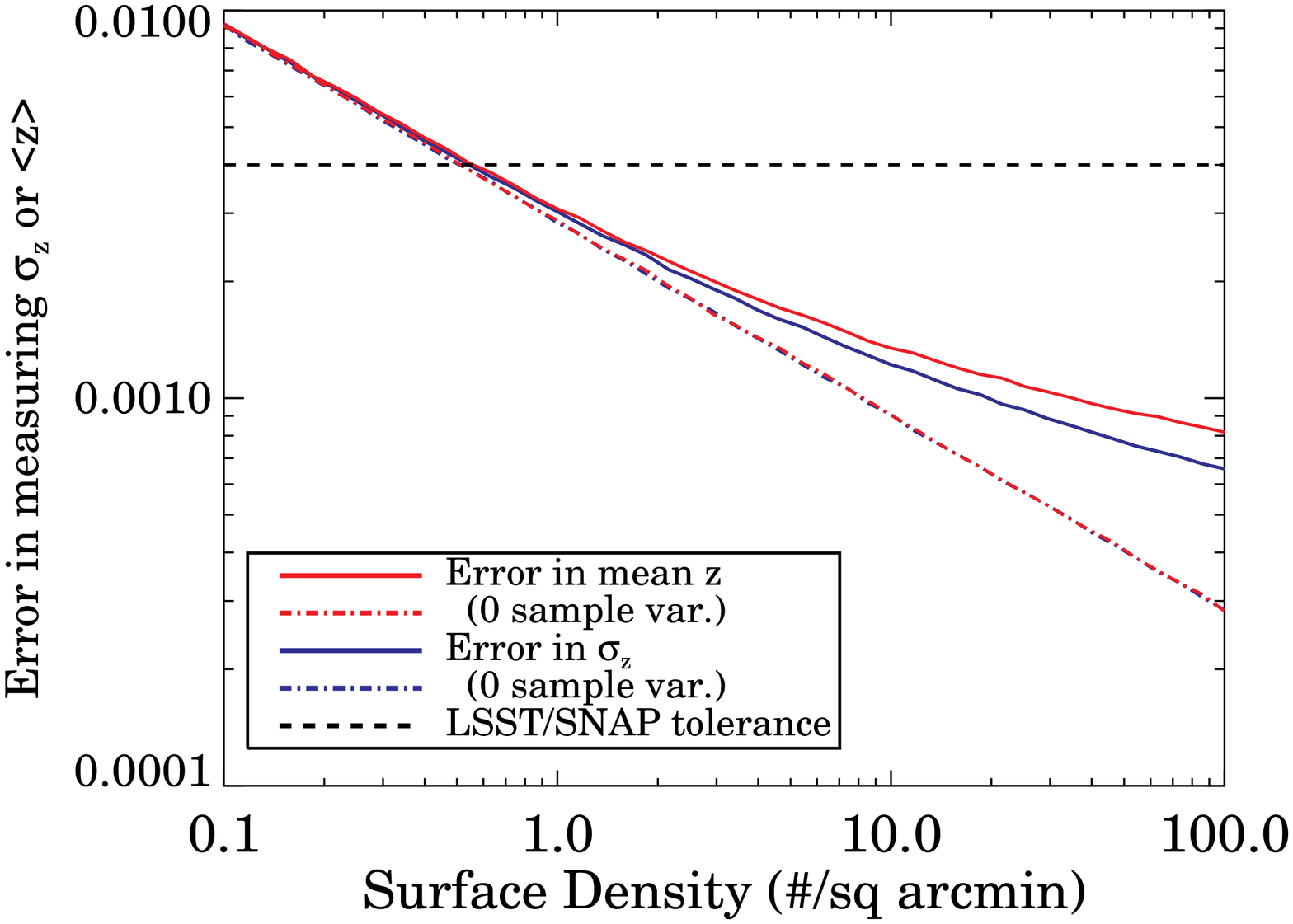}
\caption{Errors in the recovery of the mean redshift, $\langle z \rangle$ (red) or the RMS dispersion of redshifts, $\sigma_z$ (blue) for objects in a photometric sample versus their surface density in galaxies per square arcminute, $\Sigma_p$, as measured in our Monte Carlo simulations.  Note that for a single photometric redshift bin drawn from a larger sample, $\Sigma_p$ is the surface density only for objects in that bin, not for the overall sample.  The black, dashed line indicates the estimated maximum error in $\langle z \rangle$ allowable for proposed dark energy surveys using the SNAP satellite or LSST.  We assume a spectroscopic sample with $dN_s/dz = 25,000$ (roughly corresponding to current samples at $z \sim 1$) and a true \phipz\ having $\sigma_z=0.1$.  If sample variance is negligible, both errors scale as $\sigma \propto \Sigma_p^{-1/2}$; if it has maximal impact, their scaling is weaker, $\sigma \propto \Sigma_p^{-0.3}$.  If the photometric sample has very low surface density, larger numbers of redshifts or a narrower redshift distribution than assumed in our standard scenario may be required to meet the requirements of future dark energy surveys.\label{fig:sden}}
\end{figure}

\begin{figure}
\epsscale{1.20}
{\hskip -0.1in \plotone{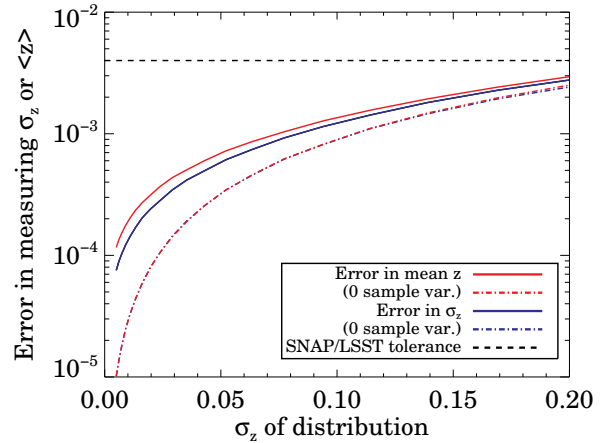}}
\caption{Errors in the recovery of $\langle z \rangle$ (red) or $\sigma_z$ (blue) versus the true value of $\sigma_z$, from our Monte Carlo tests.  The black, dashed line indicates the estimated maximum error in $\langle z \rangle$ allowable for proposed dark energy surveys using LSST or the SNAP satellite.  We assume here a spectroscopic sample with $dN_s/dz = 25,000$ (roughly corresponding to current samples at $z \sim 1$) and a photometric sample with a surface density of 10 galaxies per square arcminute.   If sample variance is negligible, both errors scale as $\sigma_z^{3/2}$; if it has maximal impact, their $\sigma_z$--dependence is weaker, $\sigma \propto \sigma_z$.  In all plotted cases, the errors in measuring the parameters of the redshift distribution are much smaller than required for future dark energy surveys.  \label{fig:sigmaz}}
\vskip 0.1in \end{figure}

\begin{figure}
\epsscale{1.20}
\plotone{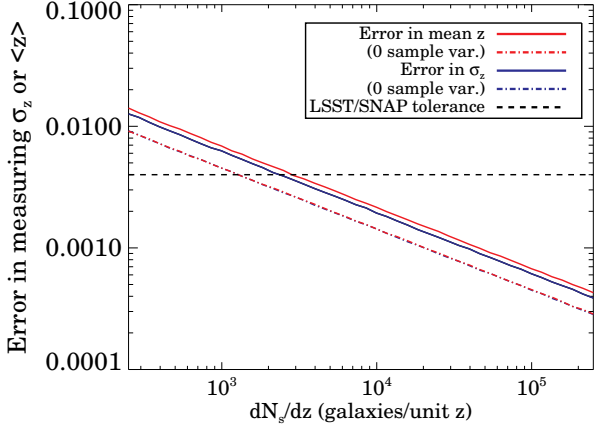}
\caption{Errors in the recovery of $\langle z \rangle$ (red) and $\sigma_z$ (blue) versus the number of spectroscopic galaxies per unit redshift, $dN_s/dz$, from our Monte Carlo tests.  The black, dashed line indicates the estimated maximum error allowable for proposed dark energy surveys using the SNAP satellite or LSST.  We assume here that the photometric sample has a surface density of 10 galaxies per square arcminute and a true \phipz\ having $\sigma_z=0.1$.   Regardless of assumptions about sample variance, all uncertainties scale as $(dN_s/dz)^{-1/2}$. If $dN_s/dz$ is small, as at $z>2$ currently, meeting the tolerances of future dark energy surveys may be problematic.  However, accuracy requirements at $z\sim 2$ are in general less restrictive than the $z \sim 1$ tolerance plotted here, as angular diameter distance and lookback time evolve more slowly with redshift at higher $z$.  \label{fig:dnsdz}}
\end{figure}

The results of these simulations are shown in Figures \ref{fig:sden}, \ref{fig:sigmaz}, and \ref{fig:dnsdz}.  For samples with constant $dN_s/dz$, if sample variance is negligible, the Monte Carlo simulations find that the errors in determining either $\langle z \rangle$ or $\sigma_z$ are identical, and can be fit extremely well (to within 1\%) by:
\begin{align}
	\sigma  = 9.1\times &10^{-4} \left({\sigma_z \over 0.1}\right)^{1.5} \left({\Sigma_p \over 10}\right)^{-1/2} \left( {dN_s/dz \over 25,000 } \right)^{-1/2} \nonumber \\
	& \times \left({4\,h^{-1}\, {\rm Mpc} \over r_{0,sp}}\right)^{\gamma} \left({10\, h^{-1} \,{\rm Mpc} \over r_{max}}\right)^{2-\gamma}  \,,
	\label{eq:scaling0cv}
\end{align}
\\
\noindent where $\Sigma_p$ is expressed in galaxies per square arcminute.  Typical values of $\gamma$ for both local and $z\sim 1$ galaxy samples are 1.7--1.8 \citep{2005ApJ...630....1Z,2006ApJ...644..671C}.

The scaling of uncertainties with $\sigma_z$ may be understood as the combination of two effects.  First, if the $x$ coordinate of a distribution is rescaled by some factor, errors in quantities proportional to $x$ should be rescaled by the same factor, so it is not surprising that $\sigma \propto (\sigma_z/0.1)$ at least.  However, there is an additional factor: when $\sigma_z$ is smaller, \phipz\ is more concentrated about the mean value, so fractional errors in \phip\ from Poisson statistics are smaller about the peak, leading to the additional factor of $(\sigma_z/0.1)^{0.5}$.



If sample variance is corrected for using the observed fluctuations in $dN_s/dz$, the uncertainty in determining $\langle z \rangle$ is fit fairly well (to within 20\%) by:
\begin{align}
	\sigma = 1.4\times & 10^{-3} \left({\sigma_z \over 0.1}\right) \left({\Sigma_p \over 10}\right)^{-0.3} \left( {dN_s/dz \over 25,000 }\right)^{-1/2} \nonumber \\
	& \times \left({4\,h^{-1}\, {\rm Mpc} \over r_{0,sp}}\right)^{\gamma} \left({10\, h^{-1} \,{\rm Mpc} \over r_{max}}\right)^{2-\gamma}  \,,
	\label{eq:scalingcv}
\end{align}
while the uncertainty in $\sigma_z$ proves to be 10\% smaller.  The dependence of errors upon $\sigma_z$ and $\Sigma_p$ is significantly weaker in this scenario.  When $\sigma_z$ is smaller, the true redshift distribution covers a smaller range in $z$, making the impact of sample variance larger; while when $\Sigma_p$ increases, Poisson errors decrease but sample variance does not, reducing its effects.

For comparison, the estimated requirements for ambitious future surveys such as SNAP or LSST are $\sigma(\langle z \rangle) \ltsim$2-4$\times 10^{-3}$ at $z\sim 1$ (see \S 1); throughout the remainder of this paper, we will take $3\times 10^{-3}$ as a reasonable target for these projects, and indicate it by a dashed line in the relevant figures.  Even with one-tenth the surface density assumed in our standard scenario, estimated errors are within this limit.    Cross-correlation techniques can meet the calibration requirements of next-generation dark energy surveys.


\subsubsection{Combining Survey Samples}

We now consider more realistic scenarios, where the spectroscopic sample is a combination of real or planned redshift surveys.  In general, different redshift surveys will be optimized for different depths or redshift ranges; hence, to cover the full possible redshift range of the photometric objects, a combination of different redshift surveys is likely to be used.  This does not present any fundamental problems;  $\xi_{sp}$ and $\xi_{ss}$ may be determined and \phipz\ estimated separately for each dataset; the resulting \phipz\ from each sample may be combined with weighted means.  

If random errors on each sample's clustering measurements are small ($\ltsim 1\%$), as is generally true in large modern surveys, the net errors on \phipz\ should be the same when we simply use the aggregate $dN_s/dz$ as if we measure separately and combine (we test the impact of clustering measurement uncertainties in \S \ref{sec:xiss}).  We hence estimate the combined $dN_s/dz$  from current samples and from surveys that have recently begun observations.  The samples considered, their estimated median redshifts, and each sample's total number of galaxies are given in Table \ref{tab:dndz}.  Where other information is not available, we have used announced or published magnitude limits and the fitting formulae from \citet{2004ApJ...617..765C} to estimate median redshifts.  We assume that except for hard redshift limits set by color cuts or lack of features in spectral windows, all $z < 2$ samples have redshift distributions of the form $z^2 e^{-z/z_0}$, which fits current datasets well \citep{2004ApJ...617..765C}; for distributions of this form, $z_0 = {\rm median}(z) / 2.67$.  For $z \ge 2$, we assume that $dN_s/dz$ will be flat, the rough consequence of applying a wide variety of high-redshift selections targeted at different $z$ ranges.  The estimated combined redshift distributions for current and near-future samples used here are shown in Figure \ref{fig:dndz}.

For the sample of current surveys (all but zCOSMOS, WiggleZ, and PRIMUS from Table \ref{tab:dndz}) we then find:           
\begin{equation}
	\sigma = 1.2\times 10^{-3} \left(\frac{\sigma_z}{0.1}\right) \left(\frac{\Sigma_p}{10}\right)^{-0.3} \,,
\end{equation}
while for a reasonable near-future scenario (including only projects that have begun observations), we find:
\begin{equation}
	\sigma = 4.1\times 10^{-4} \left(\frac{\sigma_z}{0.1}\right)^{1.5} \left(\frac{\Sigma_p}{10} \right)^{-1/2} \,,
\end{equation}
corresponding well to the tolerances for future dark energy experiments ($\sigma(\langle z \rangle) \ltsim 3\times 10^{-3}$, as described in \S 1).  We have assumed that sample variance is negligible for the near-future scenario, as the largest-area projects (SDSS and WiggleZ) plan to survey at least 1000 square degrees each (see \ref{sec:cv}).

\subsection{Non-parametric reconstruction}

Although these methods are highly successful at reconstructing the parameters of \phipz\ given the correct general model, we can also test how well we may recover $\langle z \rangle$ when making no assumptions about \phipz\ at all.  
We hence calculate the recovered mean redshift for each Monte Carlo realization, $\langle z\rangle = \sum z_i \phi_i / \sum \phi_i$, where $z_i$ is the redshift of the $i$th bin, $\phi_i$ is the recovered \phipz\ in that bin, and we use $\sum$ to indicate summation over all bins $i$.  We then find that, for our standard scenario and averaging over the redshift range $0<z<2$, the standard deviation of these mean redshift estimates is:
\\
\begin{align}
	\sigma(\langle z\rangle) = 6.9 \times & 10^{-3}  \left({\Sigma_p \over 10}\right)^{-1/2} \left( {dN_s/dz \over 25,000 }\right)^{-1/2} \nonumber \\
	& \times \left({4\,h^{-1}\, {\rm Mpc} \over r_{0}}\right)^{\gamma}  \left({10\, h^{-1} \,{\rm Mpc} \over r_{max}}\right)^{2-\gamma}  
	\label{eq:nperror}
\end{align}
\vskip 0.002in \noindent if sample variance is corrected using the observed $dN_s/dz$; the prefactor is $6.4 \times 10^{-3}$ if sample variance is negligible.  These errors would be reduced if the redshift range considered is more limited.  We plan to further explore the effectiveness of nonparametric reconstruction of redshift distributions in future work.  



\subsection{Impact of redshift outliers}

The objects that fall within a photometric redshift bin generally are not a pure population.  Non-Gaussian photometric errors, e.g. due to contamination by light from overlapping objects, may cause some galaxies to incorrectly be placed in a given bin (our sample $p$), while in other cases the observed colors of galaxies at very different redshifts may be degenerate, causing \phipz\ to be multimodal.  

\begin{figure}
\epsscale{1.20}
\plotone{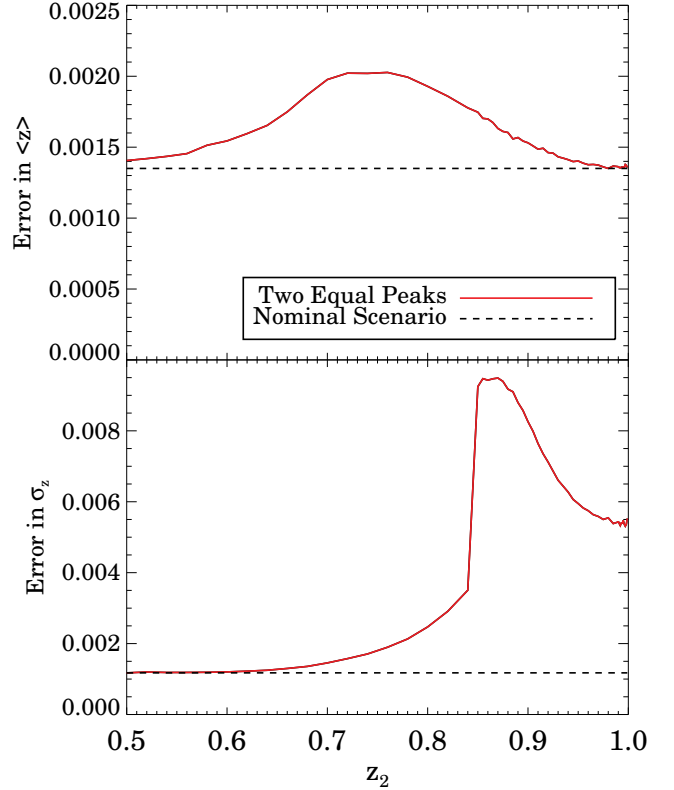}
\caption{Errors in the recovery of the mean redshift of the photometric sample, $\langle z \rangle$ (top), or the Gaussian $\sigma$ of the redshift distribution, $\sigma_z$ (bottom), for scenarios where the redshift distribution of the photometric sample \phipz consists of two Gaussian peaks of equal height, one centered at redshift 1 and the other at redshift $z_2$.  We assume for these simulations that the photometric sample has a surface density of 10 galaxies per square arcminute, that each peak has Gaussian $\sigma_z=0.1$, and that the spectroscopic sample has $dN_s/dz = 25,000$.  Although recovery of $\sigma_z$ is degraded when the two peaks are not resolved, measurement of $\langle z \rangle$ improves compared to the intermediate regime.  Results are qualitatively similar if the true $\sigma_z$ is changed. \label{fig:2peaks}}
\end{figure}

If \phipz\ consists of a combination of multiple Gaussians that overlap only minimally, the scalings from Equation \ref{eq:scaling0cv} should hold for each peak, save that we must replace $\Sigma_p$, the total surface density of photometric objects, by $f \Sigma_p$, where $f$ is the fraction of sample $p$ that is associated with a given peak.  As the peaks begin to overlap, however, this prescription will fail.  We have therefore adapted our Monte Carlo simulations to test the recovery of a distribution function consisting of two Gaussian peaks of equal width and amplitude (centered at redshifts $z_1$ and $z_2$) as $z_2$ approaches $z_1$; i.e., we employ a distribution function \phipz$= 1/(8 \pi \sigma_z^2)^{-1/2} (\,{\rm exp}( -(z-z_1)^2/2\sigma_z^2)+{\rm exp}( -(z-z_2)^2/2\sigma_z^2)\,)$.  For convenience, we take $z_1=1$ and $z_2 < z_1$, though behavior should not depend strongly on these choices.  The uncertainties in measuring $\langle z \rangle$ and $\sigma_z$  for $\sigma_z=0.1$ are shown in Fig. \ref{fig:2peaks}; we obtain qualitatively similar results for $\sigma_z=0.05$ or 0.2.

Matching the predictions above, when the two peaks overlap minimally, the error in the recovered mean, $\langle z \rangle=(\langle z_1\rangle+ \langle z_2\rangle)/2$, is equal to the sum in quadrature of the errors in determining each peak's position, divided by two.  This is equal to the error obtained for a single Gaussian peak (since $f=0.5$, so the errors in $\langle z_1\rangle$ and  $\langle z_2\rangle$ are $\sqrt{2}$ times larger than the single-peak error, but the error of the mean of the two quantities is $1/\sqrt{2}$ as large as the error in one)
As $z_2$ approaches 1, errors reach a maximum of $\sim 1.5\times$ the minimum value when $(z_2-z_1)\approx 2.5\sigma_z$, and then decrease monotonically, approaching the minimum value again when $(z_2-z_1)<<\sigma_z$.  The behavior of the uncertainty in recovering $\sigma_z$ is more complex, rising rapidly (by $>5\times$) when the two peaks are unresolved ($(z_2-z_1)\ltsim 1.2 \sigma_z$); fortunately, dark energy experiments are generally less affected by errors in $\sigma_z$ than $\langle z \rangle$ \citep{2006ApJ...636...21M}.  

Alternatively, we can consider a scenario where redshift outliers share the same mean as the overall sample, but have a broader $\sigma_z$.  
For this, we consider a distribution function \phipz$=(1-f_{outlier})(2 \pi \sigma_1^2)^{-1/2} \,{\rm exp}( -(z-z_0)^2/2\sigma_1^2)+f_{outlier} (2 \pi \sigma_2^2)^{-1/2}{\rm exp}( -(z-z_0)^2/2\sigma_2^2)\,)$; that is, the sum of two Gaussians, with total probability of (1-$f_{outlier}$) and $f_{outlier}$ and standard deviations $\sigma_1$ and $\sigma_2$, respectively, but sharing the same mean, $z_0$.  As before, we produce Monte Carlo realizations for the recovery of such a distribution with cross-correlation techniques, and then fit for $f_{outlier}$, $z_0$, $\sigma_1$, and $\sigma_2$.  For an initial guess for each realization, we take random values of $z_0$, $\sigma_1$, and $\sigma_2$ with an RMS dispersion of 10\% about their true value, and a value of $f_{outlier}$ with RMS dispersion 20\% of its true value.  

\begin{figure}
\epsscale{1.2}
{\hskip -0.15in \plotone{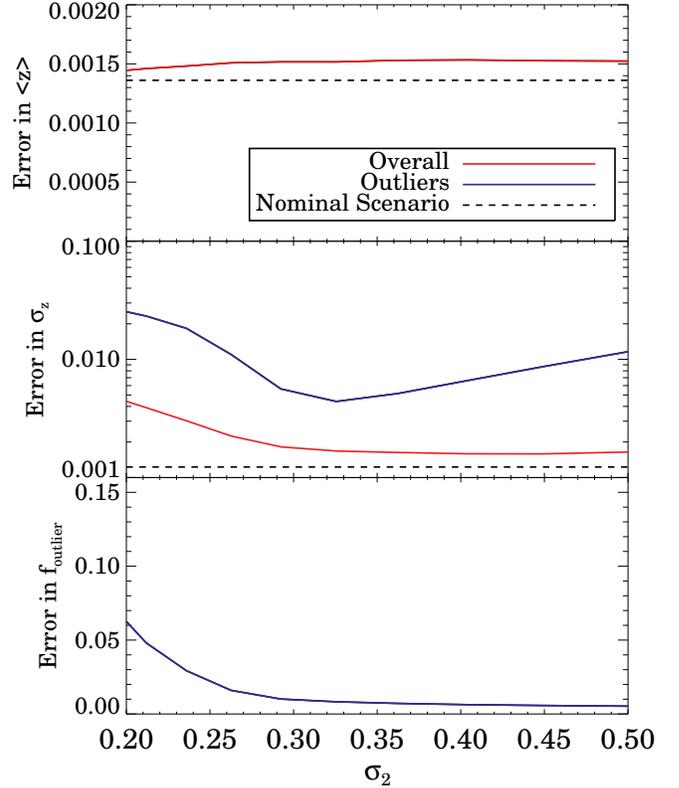}}
\caption{
Errors in the recovery of the mean redshift of the photometric sample, $\langle z \rangle$ (top), the Gaussian $\sigma$ of a redshift distribution, $\sigma$ (middle), or the outlier fraction $f_{outlier}$ (bottom) for scenarios where the redshift distribution of the photometric sample \phipz consists of two Gaussian peaks with equal mean, one having integral (1-$f_{outlier}$) and $\sigma=0.1$, and the other having integral $f_{outlier}$ and $\sigma=\sigma_2$.  All curves are plotted as a function of the Gaussian $\sigma$ of the outlier redshift distribution, $\sigma_2$.  We assume for these simulations that the photometric sample has a surface density of 10 galaxies per square arcminute and that the spectroscopic sample has $dN_s/dz = 25,000$.  Red curves indicate the error in the overall mean or net $\sigma$ of the distribution; blue curves indicate the error in recovering the width of the outlier distribution, $\sigma_2$ or the outlier fracton.  The dashed line in each case shows what the errors would be if $f_{outlier}=0$.  In the top panel, only one curve is shown, as the two Gaussians are required to have equal mean while fitting.  Results are qualitatively similar if the value of $\sigma_1$ is 0.05 or 0.2. \label{fig:outliers}}
\end{figure}

For convenience, we simulate distributions $z_0 =1$ and $f_{outlier}=0.1$, a realistic value for faint samples (see, e.g., Ilbert et al. 2006).  We set $\sigma_1$ = 0.05, 0.1, or 0.2, and investigate the recovery of $\langle z \rangle$, $\sigma_2$, and the net $\sigma$ of the distribution, 
$(\,(1-f_{outlier})^2\sigma_1^2+f_{outlier}^2\sigma_2^2)^{1/2}$, as a function of $\sigma_2$.  Results for $\sigma_1=0.1$ are shown in Fig. \ref{fig:outliers}.

The conclusions are similar for all three values of $\sigma_1$.  As seen in the top panel of Fig. \ref{fig:outliers}, the presence of a small fraction of objects with a greater $\sigma_z$ causes a correspondingly small (10-20\%) degradation in the recovery of $\langle z \rangle$.  This is not be a great surprise.  Since the two Gaussians are required to have the same mean, the principal impact on $\langle z \rangle$ is due to the broader effective $\sigma$ of the distribution; as we found before, reconstruction of $\langle z \rangle$ grows poorer when $\sigma_z$ is greater. 

When $\sigma_2$ has a value close to that of $\sigma_1$, differentiating the two components becomes difficult; therefore, errors in both $\sigma_1$ and $\sigma_2$ are greater for low values of $\sigma_2$.  Errors in $f_{outlier}$, too, are correspondingly higher in this regime.  The result is that the net $\sigma$ of the distribution can be degraded substantially where $\sigma_2 \approx 2 \sigma_1$, by more than a factor of 3.  For smaller values of $\sigma_2$ than this, analysis becomes difficult, as the fitting routine will trade off which Gaussian component corresponds to which piece of the distribution.  As $\sigma_2$ increases past the point where the two distributions become distinguishable, the error in recovering $\sigma_2$ also steadily increases, consistent with the increase in errors in $\sigma_z$ as $\sigma_z$ increases for our standard scenario.  However, $f_{outlier}$ is better determined as $\sigma_2$ goes up, such that the RMS error in the net $\sigma$ decreases monotonically as $\sigma_2$ increases.

%

\subsection{Impact of sample variance}

\label{sec:cv}

In \S \ref{sec:results}, we established the minimum and maximum impact sample variance will have on measurements using cross-correlation techniques.  Here, we attempt to establish quantitatively in which regimes a correction using the observed $dN_s/dz$, as described in \S \ref{sec:cvtheory}, will mitigate the additional errors caused by sample variance, and in which regimes cosmic variance is negligible and such a correction is inadvisable.  

To do so, we have performed another set of Monte Carlo simulations, in which we have added in quadrature to the result of Equation \ref{eq:errors} an additional uncertainty corresponding to the fluctuations in the count of an unbiased tracer of dark matter in each bin due to sample variance when producing each realization.  We assume that a total of $N_{fields}$ independent (i.e., widely separated) fields are covered by the spectroscopic samples, with each field having the same dimensions on the sky.  For simplicity, we consider only two field geometries here: either $1 \deg \times 0.5 \deg$ (correspondingly roughly to the sizes of the independent fields surveyed by current deep surveys such as DEEP2 and the VIMOS-VLT Deep Survey [VVDS]), or $2 \deg \times 2 \deg$ (corresponding to proposed future surveys).  We calculate the uncertainties from sample variance using the methods of \citet{2002ApJ...564..567N}, and find that the expected fractional root-mean-square (RMS) variations in counts of an unbiased tracer are 48\% for the smaller field size and 22\% for the larger over $\Delta z$=0.01.\footnote{IDL code is available at {\tt http://astro.berkeley.edu/ \~{ }jnewman/research.html}. } Although the fields are $8\times$ larger in area for $2 \deg \times 2 \deg$ fields, the fluctuations in counts due to sample variance are only $\sim2.2\times$ smaller; the power spectrum remains non-negligible on degree scales at $z\sim 1$.  We use the same parameters as for our standard scenario here ($\Sigma_p=10$ galaxies/square arcminute, $dN_s/dz=25,000$, and $\sigma_z=0.1$).  


For these calculations, we treat the fluctuations in counts from sample variance in successive redshift bins as independent; this assumption is fairly good (e.g. for $1 \deg \times 0.5 \deg$ field sizes and $\Delta z$=0.01, the covariance between counts in adjoining redshift bins is roughly 16\% of the total variance; for $\Delta z$=0.05, it is only 3\%).  The root-mean-square errors in recovering $\langle z \rangle$ as a function of $N_{fields}$ are shown in Figure \ref{fig:sv}.  For $1 \deg \times 0.5 \deg$ fields, the uncertainty when sample variance fluctuations are not corrected for is worse than the errors resulting from using the observed $dN_s/dz$ to make corrections so long as $N_{fields} \ltsim 15$.  For the larger field size, sample variance is worse than the uncertainties in the correction only for $N_{fields} \ltsim 3$.  If more fields than this are surveyed, there is no advantage to using fluctuations in the spectroscopic redshift distribution to correct for sample variance.  A number of fields roughly $4-5\times$ greater is required for sample variance to have completely negligible impact.  

\subsection{Covariance of Sample Variance}

\label{sec:cvcovariance}

Our Monte Carlo simulations assumed that fluctuations in counts due to sample variance are independent between all redshift bins.  However, this is of course not the case; e.g., high peaks will tend to cluster together.  We therefore must assess to what degree this covariance will worsen the reconstruction of \phipz.  

A simple way to test this is to determine how errors in the parameters of \phipz\ change when the redshift bin size used is altered.  The Monte Carlo simulations described above incorporate the total sample variance within whatever bin size is used; e.g., the errors from sample variance used if bin sizes double are not simply $1/\sqrt{2}$ as large as before.  One caveat for this test is that recovery of the parameters of \phipz\ may be affected by discreteness effects (the mean of \phipz\ within a bin is not identical to the value of \phipz\ at the bin's center, although it is treated as such in fitting) when redshift bins become large, independent of any sample variance effects.

For our standard scenario ($\sigma_z=0.1$, 4 independent fields, $\Sigma_p=10$), we find that errors in $\langle z \rangle$ and $\sigma_z$ rise steadily as bin size increases if the observed $dN_s/dz$ is not used to correct for sample variance fluctuations, reaching 25\% larger values for $\Delta z=0.1$ bins than for $\Delta z=0.01$.  If $dN_s/dz$ is used for corrections, however, errors in reconstruction are flat with bin size to 1\% or better; this is no surprise, as in such scenarios, we are limited by Poisson uncertainties in the correction (which should have no covariance between bins) rather than the sample variance itself.

Because of the possibility of discreteness effects, we have investigated the effects of the covariance of sample variance with a model that may be employed even for small bin sizes.  We proceed by showing that the dominant effect is a covariance only between successive redshift bins, with larger-scale effects being comparatively negligible; and then show that incorporating this leading-order effect causes only a modest degradation in the recovery of $\langle z \rangle$ and $\sigma_z$.  

Specifically, let us suppose that the fractional fluctuation in a count from sample variance in the $i$th redshift bin, which we will label $s_i$, is covariant only with the fluctuations in the neighboring bins ($s_{i-1}$ and $s_{i+1}$.  We also assume that the RMS variations from sample variance in each bin are equal  -- this holds to high accuracy to high accuracy for bins of constant $\Delta z$ (based upon tests with the QUICKCV code from Newman \& Davis 2002) -- and that the covariance between bins similarly does not depend on $z$; we only need these assumptions to hold locally (i.e., for small $\Delta_i$).  Even if present, small asymmetries in the impact of sample variance with redshift would affect this test only modestly, however, as they would simply mean that the RMS impact is slightly less on one side than another, but the overall effect of the covariance would remain largely unchanged.  If the impact of covariance proved to be large, this might begin to make a quantitative difference, and it might be necessary to include such effects.  

As an additional caveat, because it calculates in real space, QUICKCV does not account for the fact that correlation functions and the power spectrum are asymmetric in redshift space.  Peculiar velocities will cause the true covariance between successive bins to be overestimated by the procedure below, as their dominant effect on large scales is the "Kaiser infall" \citep{1987MNRAS.227....1K}, which causes large structures to appear collapsed along the line-line-of-sight in redshift space.  This is particularly the case for optically-selected samples at high redshift, which are biased towards intrinsically blue galaxies and have only very weak "Fingers of G-d" \citep{2007arXiv0708.0004C}; or if the most nonlinear, sub-Mpc scales are excluded from cross-correlation analyses, as suggested in \S \ref{sec:errors}.  The net effect of redshift-space distortions is that sample variance fluctuations will be more confined to a single redshift bin than one would expect from a real-space calculation, so our estimated errors from this model will be conservative (i.e., biased high).

Given our assumptions, we may presume that for each bin there is an underlying, 'hidden' variable, $s'_i$, which has Gaussian random variations completely independent of the adjoining bins; and we can write
\begin{equation}
\label{eq:cvmodel}
s_i = (1-2w)s'_i + w s'_{i-1} + w s'_{i+1} ,
\end{equation}
where $w$ is some unknown weight factor.  We take the RMS variation for each of the uncorrelated $s'_i$ to be given by the variable $\sigma_u$ (for uncorrelated), which by assumption is the same for all $i$.   

Given this model and standard propagation of errors, it is possible to predict the net fluctuation due to sample variance for a single bin of width $\Delta z$, $2 \Delta z$, $3 \Delta z$, etc.  The first three of these are:
\begin{eqnarray}
& \sigma_{\Delta z}^2 & = (6 w^2 - 4 w +1) \sigma_u^2  \\
& \sigma_{2\Delta z}^2 & = {1 \over 4} (4 w^2 - 4 w +2) \sigma_u^2  \\
& \sigma_{3\Delta z}^2 & = {1 \over 9} (4 w^2 - 4 w +3) \sigma_u^2 .
\end{eqnarray}
Therefore, given numerical predictions for $\sigma_{\Delta z}$ and $\sigma_{2\Delta z}$ given the full power spectrum (predictions which we take from QUICKCV), it is possible to solve for $w$ and $\sigma_u$ in this model.  We can then use the same code to determine $\sigma_{3\Delta z}$, and compare it to the prediction of the model, in order to assess the model's effectiveness at incorporating the impact of the covariance of sample variance.  

We find that this simple model is highly effective.  For bin sizes (i.e. $\Delta z$) ranging from 0.003 to 0.1, ignoring the covariance of sample variance completely (so that $ \sigma_{3\Delta z} = \sigma_{\Delta z} / \sqrt{3}$) underpredicts the RMS fractional variation in a count in a bin of width $3\Delta z$ by anywhere from 2\% (for $\Delta z=0.1$) to 23\% (for $\Delta z=0.003$); for our standard $\Delta z=0.01$ bins, the underprediction is 15\% (all tests are for a 1 deg $\times$ 1 deg field with central redshift $z=1$).  If we employ the model described above, however, the underprediction ranges from below 0.1\% (for $\Delta z \ge 0.025$) to 3\% (for $\Delta z=0.003$); for $\Delta z=0.01$ the prediction is off by 0.6\%.  Clearly, the vast majority of the effect of the covariance of sample variance is described simply by a covariance between successive redshift bins.  This holds true even if we had tested over a larger $z$ range; e.g., for $\Delta z=0.01$, the model correctly predicts the net variance over 5$\Delta z$ to within 1.5\%, or over 10$\Delta z$ to within 2.1\%.  However, in the latter case, the assumption that all the bins are statistically independent performs even better, matching to 0.4\%; so this model will actually overpredict the impact of the covariance of sample variance on large scales.

\begin{figure}
\epsscale{1.2}
{\hskip -0.15in \plotone{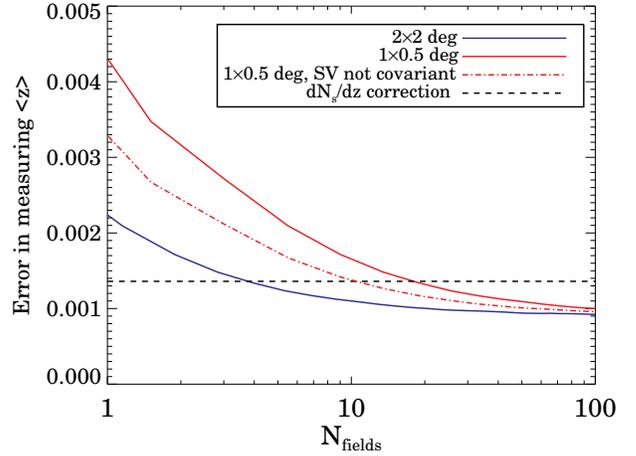}}
\caption{Errors in the recovery of $\langle z \rangle$ versus the number of independent fields surveyed for two different field geometries (red and blue curves), resulting from Monte Carlo tests in which sample variance errors are added to our standard scenario ($\Sigma_p=10$ galaxies per square arcminute, $\sigma_z=0.1$, $dN_s/dz = 25,000$).  As the number of fields increases or if larger fields are used, errors from sample variance decrease.  However, an upper limit on the practical impact of sample variance is set by the black, dashed curve, which indicates the errors if the observed $dN_s/dz$ distribution is used to correct for the density fluctuations in each redshift bin.  The errors after this correction are set by Poisson statistics from the spectroscopic sample, rather than by clustering.  For current redshift samples, applying such a correction (as assumed in the preceding plots) is favored; however, WiggleZ and subsequent surveys  should cover $\sim 1000$ square degrees each, so sample variance will affect them only minimally. The red, dot-dashed curve shows what the errors would be if sample variance were not covariant between redshift bins; see \S \ref{sec:cvcovariance}.  \label{fig:sv} }
\end{figure}

The form of equation \ref{eq:cvmodel} is particularly convenient for incorporation into our Monte Carlo tests.  Instead of simply randomly drawing the fractional fluctuation from sample variance in a given bin, $s_i$, as before, instead we can draw the uncorrelated $s'_i$, and then construct the correlated $s_i$ from the uncorrelated $s'_i$ in making each Monte Carlo realization; we need only predict $w$ and $\sigma_u$ (which we do using QUICKCV).  The impact of adding this covariance to our models may be seen in Fig. \ref{fig:sv}; errors are increased by roughly 30\% in the worst case (for $N_{fields}=1$), but by $< 15\%$ for $N_{fields} > 10$, the regime in which correcting for sample variance fluctuations with the observed $dN_s/dz$ becomes ineffective.   For our scaling scenario ($N_{fields}=4$), errors are degraded by 27\%, slightly worse than the difference between $\Delta z=0.01$ and 0.10.  We expect that any corrections from the much smaller, larger-range covariance would be considerably less than this; hence, we conclude that the covariance of sample variance has relatively minor impact on our results.  The prefactor of $9.1\times 10^{-4}$ in \ref{eq:scaling0cv} becomes $1.2\times 10^{-3}$ when this covariance is accounted for, still well within SNAP and LSST requirements.

The impact is even smaller for our standard scenario, for which the observed $dN_s/dz$ is used to correct for sample variance in each redshift bin.  In that case, the Poisson uncertainty in this correction is far greater than the covariance from sample variance, and the latter becomes totally negligible; hence, all the major results of this paper are unaffected by this covariance.

\section{Possible Systematics}

\label{sec:systematics}



We now consider the impact of a variety of effects that violate the simple assumptions underlying our basic scenario.  We will treat these errors analytically wherever possible.  We summarize the results of this section in Table \ref{tab:errors}.

\subsection{Evolution in bias}

\label{sec:bias}

Inspection of Equation \ref{eq:wsp-phip} shows that to transform the cross-correlation signal \wsp\ into \phipz, we require knowledge of the parameters of $\xi_{sp}$, namely $\gamma$ and $r_{0,sp}$, as a function of redshift.  As described in \S \ref{sec:basics}, however, $w_{pp}$ provides constraints only on the mean clustering of the photometric sample.  What is the impact on the derived \phip\ if these parameters evolve?

We test this by assuming that the net change in $\xi_{sp}$ with redshift is due only  to changes in the linear bias of of the photometric sample, $b_p$.  We further assume that the evolution of the bias is linear; i.e., $b_p(z) = b_p(1) + (db/dz) (z-1)$ with constant $db/dz$, and that $db/dz$ is small compared to $b_p(1)$.  As usual, we take \phipz\ to be given by Equation \ref{eq:gz}, and adopt the simple-biasing assumption $\xi_{sp}=(\xi_{ss} \xi_{pp})^{1/2}$, so $\xi_{sp} \propto ({b_p(z) / b_p(1)})^{1/2} $.  Then, if $b_p$ varies but $w_{sp}$ is interpreted with a constant $b_p$, to leading order in $db/dz$ the measured $\langle z \rangle$ will be off by:
\begin{align}
\Delta(\langle z \rangle) & =  {\int z \left({b_p(z) \over b_p(1)}\right)^{1/2} \,g(z) dz \over \int \left({b_p(z) \over b_p(1)}\right)^{1/2} \,g(z) dz} - {\int z \,g(z) \,dz \over \int g(z) \,dz} \nonumber \\
&\approx  {1 \over \sqrt{2 \pi} \sigma_z} \int - z {db/dz \over 2 \,b_p(1)} (z-1) \ e^{-{(z-1)^2 \over 2\sigma_z^2}} dz \nonumber \\
& =  {db/dz \over 2 \,b_p(1)} \sigma_z^2 , 
	\label{eq:bias}
\end{align}
\noindent applying the linear approximation $(1 +\epsilon)^{1/2} \approx (1+\epsilon/2)$.  

By comparing the observed galaxy clustering to the predicted clustering of dark matter in a $\sigma_8=0.9$ model from \citet{2003MNRAS.341.1311S}, \citet{2006ApJ...644..671C} estimate that the linear bias of $\sim L_*$ galaxies at $z\sim 1$ is $1.48 \pm 0.04$.   Applying the same methods at $z=0$, the correlation measurements of \citet{2005ApJ...630....1Z} correspond to a linear bias for $L_*$ galaxies in the Sloan Digital Sky Survey of 1.0, differing by $\ltsim 0.5 \pm 0.1$.  We therefore take 0.5 to be a reasonable upper limit on $[ (db/dz) /b_p(1) ]$.  If the bias of the photometric sample evolves but we do not take that into account in our analysis, we would make a systematic error of at most 0.0025 for $\sigma_z=0.1$, pushing the tolerances of future dark energy projects.  In actuality, however, we do possess information on bias evolution (e.g. by measuring the mean $b_p$ over successive photometric bins), so it is very unlikely this effect would lead to an error that large.  

It is worth noting that spectroscopic surveys with multiple selection techniques (e.g. by a variety of different color cuts) may select objects with very different bias according to the technique used. This could, for instance, lead to jumps in bias at the redshifts where a given subsample becomes relevant or irrelevant.  In that case, it would be most effective to split the survey into its constituent subsamples, treating the clustering/bias of each subsample separately.

\subsection{Errors in measuring autocorrelations of spectroscopic samples}

\label{sec:xiss}

A second potential issue is systematic errors in measuring the autocorrelation function of the spectroscopic sample, $\xi_{ss}$.  Although random errors in modern surveys are small, it is very difficult to reduce systematic errors below 1-2\%.  These systematic errors will generally cause $\xi_{ss}$ to be over- or under-estimated similarly at all $z$ covered by a given survey.  To assess the impact of these systematics, we test their worst-case impact using our standard \phipz\ distribution.  

Thus, we assume that there are only two redshift surveys; that one survey covers the complete $z \leq 1$ regime, and another $z>1$; and that the $\xi_{ss}$ measurement for each survey may suffer an unknown systematic of RMS fractional amplitude $\sigma_{sys}$; i.e., if $\sigma_{sys}=0.02$, we expect the measured amplitude of $\xi_{ss}$ (for similar galaxies at the same redshift) to vary by 2\% from survey to survey.  Thus, one half of the reconstructed \phipz\ will have an amplitude differing from the other by a factor $r_{sys}$ drawn from a distribution with RMS $\sigma_{sys}$ and mean 1 (the difference is $\sigma_{sys}$ rather than $2 \sigma_{sys}$ due to the simple-biasing assumption and linear approximation to the square root, as in \S \ref{sec:bias}).  Then:
\begin{align}
\Delta(\langle z \rangle) & =  {\left[\int_{-\infty}^1 z \,g(z)\, dz + r_{sys} \int_1^\infty z\, g(z) \,dz \right]  \over \left[\int_{-\infty}^1 g(z) \,dz + r_{sys} \int_1^\infty g(z) \,dz \right]} - 1 \nonumber \\
 &= {r_{sys}-1 \over r_{sys}+1} {\sqrt{2} \sigma_z \over \sqrt{\pi}} \nonumber	\\
 &\approx {r_{sys}-1 \over 2} {\sqrt{2} \sigma_z \over \sqrt{\pi}} ,	\label{eq:biasratio} 
\end{align}
taking $\sigma_{sys}$ to be small, so $r_{sys} \approx 1$.
Propagating errors, we then find that in a worst-case scenario, $\langle z \rangle$ will have a root-mean-square bias of $(1/ 2\pi)^{1/2}\sigma_{sys}\sigma_z$, or more conveniently, $8.0 \times 10^{-4} (\sigma_{sys}/0.02) (\sigma_z/0.1)$, well within estimated tolerances for SNAP and LSST.  

\subsection{Field-to-field zero point variations}

A third factor not considered in our standard scenario is spatial variation in the effective zero points of the photometry used to define the photometric sample, $p$ (due to seeing, calibration issues, etc.).  Note that random photometric errors or absolute zero point uncertainties have no effect on redshift distributions measured from cross-correlations; the method will empirically determine \phipz\ for whatever falls in a given photometric redshift bin, regardless of whether a given object is put in that bin due to errors or because it rightfully belongs there.  Instead, the principal impact of zero point errors will be changes in the effective depth of the sample between separately calibrated patches;  an error of $\Delta m$ magnitudes will lead to a fractional error in number counts of $N^{-1} dN/dM \Delta m$.  For $R$-band-limited samples, $d ({\rm log}_{10} N)/dM = N^{-1}/(\ln{10}) dN/dM  \approx 0.36$ \citep{2004ApJ...617..765C}, so the fractional variation in the surface density of objects in the photometric sample ($\Sigma_p$), will be approximately $0.83 \,\sigma_{zp}$, where $\sigma_{zp}$ is the RMS variation of the photometric zero point in magnitudes.  The logarithmic slope of galaxy number counts is larger in $B$ ($\sim 0.5$) and slightly smaller for $I$ ($\sim 0.33$), leading to modestly different prefactors for these cases.

Zero point variations will impact the errors in determining \phipz\ in two ways.  The first is that, if different spectroscopic surveys cover regions with different photometric zero points, the cross-correlation signal will be artificially boosted or decreased for each survey as $\sigma_{zp}$ varies, since the overall value of $\Sigma_p$ used for normalization will be not quite appropriate for the effective magnitude limit in each patch of sky.  Again, we consider a worst-case scenario, where one survey covering one set of $N_{patch}$ independent calibration patches is used to reconstruct \phipz\ at $z \leq 1$, and another covering a separate set of $N_{patch}$ patches is used for $z > 1$.  

We may again apply the results of Equation \ref{eq:biasratio}.  Since the fractional error in $w_{sp}$ for a single survey will be $0.83 \,\sigma_{zp} \,N_{patch}^{-1/2}$ (as we are averaging over $N_{patch}$ calibration patches), the RMS variation in $r_{sys}$ (the ratio of the reconstructed \phipz\ at $z>1$ to $z<1$) should be $\sqrt{2} \times 0.83 \,\sigma_{zp} \,N_{patch}^{-1/2}$.  Thus, the worst-case RMS error in a measurement of $\langle z \rangle$ due to zero point variations is $2.3\,\times\,10^{-3}\,(\sigma_{zp}/0.01)\,(N_{patch}/4)^{-1/2}\,(\sigma_z/0.1)$, within SNAP and LSST tolerances.  Specifications for zero point variations are generally smaller than this (e.g. 0.005 mag RMS zero point variation for LSST; cf. \citealt{2006SPIE.6267E..38B}), and that ongoing redshift surveys should cover many independently-calibrated patches of sky, so zero point variations are likely to have even smaller impact.   

The second effect of zero point variations will be to increase the fluctuations in counts beyond those expected for Poisson errors (as assumed in Equation \ref{eq:errors}), even if there is only one redshift survey.  However, for reasonable scenarios, this is minor; even in a very conservative scenario, with $dN_s/dz=25,000$, $\Sigma_p=10$ galaxies per square arcmin, only 3 independently calibrated patches of sky surveyed, and RMS zero point errors of 0.05 mag, errors in mean $z$ and $\sigma_z$ from Monte Carlo tests increase by a fraction of a percent of their value when this effect is added to our standard scenario.

\subsection{Errors in assumed cosmology}

\label{sec:cosmo}

As seen in Equation \ref{eq:phip}, transforming ${\tilde w }(z)$ to \phipz\ requires knowledge of the volume element,  \dvdz$ = d_A(z)^2\,dl/dz$.  Because the scale radii used are expressed in $h^{-1}$ Mpc, all scalings with the Hubble parameter cancel out; but \dvdz\ will depend on other cosmological parameters as well.  Similarly to the preceding sections, we investigate this by assessing the error in $\langle z \rangle$ that will result from assuming a mistaken cosmology.   For convenience, we consider only spatially flat, quintessence + cold dark matter models characterized by a matter density $\Omega_m$  and dark energy equation-of-state parameter $w$; these are 0.3 and -1, respectively, for our standard scenario.  

We then determine the error in $\langle z \rangle$ that occurs if the cosmology differs from the standard in one of these parameters, but  ${\tilde w }(z)$ is interpreted using the standard cosmology.  Let $V_{assumed}(z)$ and $V_{true}(z)$ denote the values of $d_A(z)^2 \,dl/dz$ for the assumed cosmological model and the true cosmology, respectively.  Then mistaken assumptions will lead to an error in $\langle z \rangle$ given by:
\\
\begin{equation}
\Delta \langle z \rangle  =  {\int_{0}^{\infty} z\, V_{assumed}(z) \,g(z)\, dz   \over \int_{0}^{\infty} V_{assumed}(z) \,g(z)\, dz} -  {\int_{0}^{\infty} z\, V_{true}(z) \,g(z)\, dz   \over \int_{0}^{\infty} V_{true}(z) \,g(z)\, dz} \, .
\end{equation}
\vskip 0.07in\noindent For $\sigma_z \ltsim 0.3$, the effect of varying $\Omega_m$ may then be approximated well by $\Delta \langle z \rangle = 4.2 \times 10^{-4} (\sigma_z/0.1)^2 (\Delta \Omega_m / 0.03)$, where we normalize to a 10\% variation in $\Omega_m$ (comparable to errors from WMAP; \citealt{2006astro.ph..3449S}).  The effects of varying $w$ are much less symmetric about our standard scenario, so we consider $w < -1$ and $w > -1$ separately.  The impact of errors in cosmology are stronger in the former case: approximately $\Delta \langle z \rangle = 7 \times 10^{-5} (\sigma_z/0.1)^{1.9} (\Delta w/0.1)$, where $\Delta w= w-1$.  For $w > -1$, $\Delta \langle z \rangle$ has a turning point at $w \approx -0.93$; the value of $\Delta \langle z \rangle$ at that turning point is approximately $1.7\times 10^{-5} (\sigma_z/0.1)^{1.9}$.  Thus, it is more important to constrain $\Omega_m$ than $w$ when using cross-correlation techniques to determine \phipz, but regardless, the cosmological uncertainties are small compared to the requirements of proposed dark energy surveys.

\begin{deluxetable*}{lc}
\tabletypesize{\scriptsize}
\tablecaption{Summary of Random and Systematic Errors}
\tablewidth{0pt}
\tablehead{\colhead{Error type} & \colhead{Corresponding uncertainty in $\langle z \rangle$} }
\startdata
Random errors & $1.5 \times 10^{-3} \,(\sigma_z/0.1)^{1} \,(\Sigma_p/10)^{-0.3} \,( (dN_s/dz) / 25,000)^{-1/2} ~\tablenotemark{a} $ \\
Random errors, sample variance negligible & $1.0 \times 10^{-3} \,(\sigma_z/0.1)^{1.5}\, (\Sigma_p/10)^{-1/2} \,( (dN_s/dz) / 25,000)^{-1/2}$ \\
Not accounting for evolution in bias & $2.5 \times 10^{-3} [ db/dz /b_p(1)] / 0.5 \, (\sigma_z/0.1)^2$ \\
Systematic errors in autocorrelation measurements &  $<8.0 \times 10^{-4} \,(\sigma_{sys}/0.02) \,(\sigma_z/0.1)$ \\
Field-to-field zero point variations & $<2.3 \times 10^{-4} \,(\sigma_{zp}/0.01) \,(N_{patch}/4)^{-1/2}  (\sigma_z/0.1)$ \\
Errors in assumed $\Omega_m$ & $4.2 \times 10^{-4} (\sigma_z/0.1)^2 \,(\Delta \Omega_m / 0.03)$ \\
Errors in assumed $w$ & $<7 \times 10^{-5}\, (\sigma_z/0.1)^{1.9} \,(\Delta w/0.1) $ \\
\enddata
\tablenotetext{a}{Throughout this table we give the surface density of the photometric sample, $\Sigma_p$, in galaxies per square arcminute.} 
\label{tab:errors}
\end{deluxetable*}

\section{Conclusions}

In this paper, we have described a new method for recovering the redshift distribution of objects in a photometric sample by measuring their angular cross-correlations with objects in redshift survey samples as a function of spectroscopic $z$.  This technique does not require that spectroscopic samples resemble the photometric sample in galaxy properties (such as luminosity) or clustering amplitude.  We have demonstrated that in realistic scenarios, the redshift distributions of photometric samples may be determined to the precision required by proposed dark energy experiments with this technique.    We conclude here by discussing what can be done to optimize future redshift survey datasets to facilitate applications of cross-correlation techniques.  

$\bullet$ First, it is apparent from Figure \ref{fig:dndz} that there are two redshift regimes that are currently much more poorly sampled than others: $0.2 \ltsim z \ltsim 0.7$, and $z>1.4$.  Concentrating future survey efforts on these regimes would be of great benefit for application of cross-correlation methods.  Efforts to cover this lower-redshift gap are already underway.  

$\bullet$ We emphasize that, although high redshift precision is not requisite for the spectroscopic sample -- e.g., ideal photometric redshifts with $\sigma_z=0.01$ would be useable for determining the true redshift distribution for a sample with $\sigma_z \gtsim 0.1$ -- it is vitally important that the {\bf purity} of the redshifts be very high.  Otherwise, redshift outliers in the spectroscopic sample will cause distortions in the recovered \phipz.  Of course, the same holds true for any photometric redshift calibration technique; if false redshifts are used to calibrate a redshift distribution, it is quite likely to be biased in some way.  

In most surveys, high-purity redshifts are only obtained for a fraction of the sample.  For instance, in the DEEP2 Galaxy Redshift Survey, roughly 80\% of successful redshifts (i.e., $\sim 55\%$ of targeted galaxies) fall within the highest purity class, which has a $\sim 0.5\%$ error rate (based on tests with the $>2000$ objects observed multiple times), while the remainder have an error rate of roughly 5\% (Faber et al. 2006, in prep.).  As another example, in the VIMOS-VLT Deep Survey, $\sim 20\%$ of galaxies targeted yield a redshift in their highest (99\%) redshift confidence category; almost all of those high-confidence objects have $z<1$ \citep{2005A&A...439..863I}.  In contrast, in the Sloan Digital Sky Survey, which samples local galaxies with much higher signal-to-noise spectra, almost all galaxies yield an accurate redshift (Schlegel et al., in prep.).  

In general, the higher signal-to-noise spectra within a sample will generally yield a greater rate of secure redshifts; i.e., the brightest galaxies will dominate samples of high-confidence redshifts.  This can be a problem for direct calibration of photometric redshifts, but has little effect on cross-correlations; the spectroscopic sample need not include high-confidence redshifts of faint galaxies, so long as a sufficient number of brighter galaxies at the same redshift are included.  This allows larger, shallower surveys to be used to calibrate redshift distributions even of very faint photometric redshift samples.

$\bullet$ Cross-correlations can be analyzed more simply if nonlinearities have minimal impact; e.g., because the spectroscopic sample has relatively weak biasing and scale-dependence in its bias.  This suggests that blue, star-forming galaxies may be more useful for this technique than red, early-type galaxies (\citealt{2005ApJ...630....1Z, 2004ApJ...609..525C}; Coil et al. 2007, in prep.).  However, red sequence galaxies can yield relatively high-precision photometric redshifts; they therefore may make up for this disadvantage by sheer abundance.  Quasars might make an attractive population to use at high redshift given their high luminosity, but the difficulty of measuring their autocorrelations with precision \citep{2006astro.ph..7454C} and the possibility that quasars may affect the evolution of nearby galaxies may make their use to measure cross-correlations over any but the largest scales difficult.  

Above all, it is important that future photometric dark energy experiments overlap with spectroscopic surveys on the sky.  Without this, cross-correlation measurements are impossible.  These measurements are not only useful for determining redshift distributions, but also will allow correlation functions to be measured down to much lower luminosities than can be reached spectroscopically.  Furthermore, the availability of more photometry in fields with spectroscopy can improve our understanding of the spectroscopic samples by broadening spectral energy distribution measurements; having photometry in five bands for each galaxy in the SDSS spectroscopic sample has been a great boon to studies of galaxy properties.  The synergies between photometry and spectroscopy are great, and determination of redshift distributions from cross-correlations is only one of many applications, though a vital one.


\bigskip

\acknowledgments

I wish to thank Tony Tyson and Andy Connolly for their assistance with developing the portion of an LSST white paper that first described this work.  I also thank Gary Bernstein for suggesting the possibility that bias evolution could impact this method; Sarah Bridle for inquiring about errors in autocorrelations and dependences on cosmological parameters; Lloyd Knox for pressing the issue of sample variance; Nikhil Padmanabhan for inquiring about pathological cross-correlation scenarios; Charlie Conroy for assistance with bias measurements; Bhuvnesh Jain, Eric Linder, Ryan Scranton, David Wittman, and Hu Zhan for helpful discussions; the anonymous referee for useful comments; and especially Alison Coil for providing a wide variety of useful information.  I also appreciate the assistance of Tony Tyson and especially Brian Gerke for reading drafts of this paper and providing a wide variety of helpful suggestions.  This work was supported by NASA through Hubble Fellowship grant HST-HF-01165.01-A  awarded by the Space Telescope Science Institute, which is operated by AURA Inc.\ under NASA contract NAS 5-26555. I wish to also thank the staff of the Institure for Nuclear and Particle Astrophysics at Lawrence Berkeley National Laboratory for all of their invaluable assistance during my time there.

\bibliographystyle{apj}
\bibliography{xcorr}
%

\clearpage


%

\clearpage

 \end{document}